\documentclass[twocolumn,english,aps,pra,showpacs]{revtex4}

\usepackage{amsmath,graphicx,amssymb,epsfig,babel,dsfont,color,subfigure}
\usepackage{changes}	

\renewcommand{\Im}{{\rm Im}}
\newcommand{\Tr}{{\rm Tr}}
\newcommand{\rd}{{\rm d}}

\newcommand{\kb}{k_{\rm B}}

\newcommand{\re}{{\rm e}}
\newcommand{\rs}{{\rm s}}
\newcommand{\rp}{{\rm p}}
\newcommand{\ri}{{\rm i}}

\begin{document}

\title{Near-field heat transfer between {multilayer} hyperbolic {meta}materials}

\author{Svend-Age Biehs$^{1}$,  Philippe Ben-Abdallah$^{2,3}$}

\affiliation{$^1$ Institut f\"{u}r Physik, Carl von Ossietzky Universit\"{a}t, D-26111 Oldenburg, Germany.}
\affiliation{$^2$ Laboratoire Charles Fabry,UMR 8501, Institut d'Optique, CNRS, Universit\'{e} Paris-Sud 11,
2, Avenue Augustin Fresnel, 91127 Palaiseau Cedex, France}
\affiliation{$^3$ Universit\'{e} de Sherbrooke, Department of Mechanical Engineering, Sherbrooke, PQ J1K 2R1, Canada.}

\email{s.age.biehs@uni-oldenburg.de, pba@institutoptique.fr}

\date{\today}

\pacs{}

\begin{abstract}
We review the near-field radiative heat flux between hyperbolic materials focusing on multilayer hyperbolic meta-materials. 
We discuss the formation of the hyperbolic bands, the impact of ordering of the multilayer slabs, as well as the impact of 
the first single layer on the heat transfer. Furthermore, we compare the contribution of surface modes to that of hyperbolic 
modes. Finally, we also compare the exact results with predictions from effective medium theory.
\end{abstract}

\maketitle

%
%
%

\section{Introduction}

Hyperbolic materials (HM)~\cite{HuChui2002,Smith2003} have attracted a lot of attention in the last years due to their unique properties.
It could be shown that they allow for a broadband enhanced LDOS~\cite{SmolyaninovNarimanov2010}, broadband 
enhanced spontaneous emission~\cite{ZubinEtAl2012,PoddubnyEtAl2011,ChebykinEtAl2012,KidwaiEtAl2012,IorshEtAl2012,PotemkinEtAl2012,KidwaiEtAl2011,OrlovEtAl2011,KimEtAl2012,KrishnamorthyEtAl2012,GalfskyEtAl2015,WangEtAl2015}, hyperbolic lensing~\cite{JacobEtAl2006,FengAndElson2006,CegliaEtAl2014,Ben2012,Conteno2015}, negative refraction~\cite{SmithEtAl2004,HoffmanEtAl2007}, super absorption~\cite{Cuclu2012}, enhanced F\"{o}rster 
energy transfer~\cite{Cortes2013,Newman2015,BiehsMenonAgarwal2016}, and self-induced torques~\cite{GinzburgEtAl2013}. HM can be artificially fabricated by a periodic layout of 
sub-wavelength metal and dielectric components for applications in the visible. On the other hand, for 
applications in the infrared it is more beneficial to combine phonon-polaritonic/semi-conductor and dielectric components~\cite{Biehs2012,GuoEtAl2012,Biehs2,GuoJacob2013,ShiEtAl2015,HoffmanEtAl2007}. Besides such hyperbolic meta-material (HMM) structures there exist also quite a number of 
natural HM~\cite{ThompsonEtAl1998,CaldwellEtAl2014,EsslingerEtAl2014,NarimanovKildishev2015,KorzebEtAl2015}
for applications in the visible and infrared.

Here, we are mainly interested in the application of HM in the context of thermal radiation in
the near-field regime~\cite{Volokitin2007,SurfaceScienceReports,BiehsReview,ZhangReview2013,SongReview}. It is well known that in 
this distance regime the radiative heat flux can be much larger than that of a blackbody due to the extra-contribution of evanescent waves
such as surface phonon polaritons~\cite{Volokitin2007,SurfaceScienceReports,ZhangReview2013,BiehsReview,SongReview}, for instance. This super-Planckian effect has been 
measured by many different experiments using different geometries and materials~\cite{KittelEtAl2005,HuEtAl2008,ShenEtAl2008,NatureEmmanuel,Ottens2011,Kralik2012,ShenEtAl2012,GelaisEtAl2014,SongEtAl2015,KimEtAl2015,GelaisEtAl2016,LimEtAl2016,KloppstechEtAl2016}, in the last ten years. So far, there is only one experiment which has measured the heat flux of a HM~\cite{ShiEtAl2015},
although HM seem to have quite interesting features for near-field thermal radiation. For example, it 
was shown that HM could serve as an analogue of a blackbody for near-field thermal radiation~\cite{Biehs2012},
allowing for broadband thermal radiation~\cite{Biehs2012,GuoEtAl2012}. Furthermore, it has been shown that HM can
be used for near-field thermophotovoltaic applications~\cite{Nefedov2011,SimovskiEtAl2013}, that HM have a large
penetration depth of thermal radiation~\cite{Slawa2014,Tschikin2015} which is an advantage also for the transport 
of near-field thermal radiation over far-field distances~\cite{MessinaEtAl2016}. Also the possibility of having
larger conduction by thermal radiation inside a HM than conduction by phonons and electrons has been discussed~\cite{Narimanov2014}
as well as the thermodynamical potentials, the general laws of thermal radiation inside HM~\cite{Biehs2015} and 
the coherence properties in the vicinity of HM~\cite{GuoJacob2014}. Recently, also tunable hyperbolic thermal emitters
have been introduced~\cite{MoncaEt2015} and it was proposed to make use of the hyperbolic 2D plasmons in graphene ribbons 
for elevated near-field heat fluxes~\cite{LiuZhang2015}.

In this article we review the near-field radiative heat flux between HM
by focusing on multilayer hyperbolic meta-materials (mHMM), because these structures can easily
be fabricated and the numerical exact treatment is relatively simple in this case. Many
researchers have already studied the radiative heat flux of layered materials in different contexts.
For example, the impact of dissipation of thin layers was discussed~\cite{Volokitin2001} as well as
the impact of surface-polariton coupling inside thin metal~\cite{Biehs2007} and later also dielectric~\cite{FrancoeurAPL,BPAEtAl2009} 
coatings. Real multilayer structures where discussed in the context of heat radiation
between two photonic crystals~\cite{Arvind2005,TschikinPC2012} and inside of photonic crystals~\cite{TschikinPC2012,LauEtAl2008,LauEtAl2009}. The contribution
of surface Bloch modes was quantified~\cite{Pryamikov2010,Pryamikov2011} and the S-matrix and impedance method where introduced
for near-field heat transfer calculations~\cite{Francoeur2009,MaslovskiEtAl2013}. In the following we will in particular make use
of the S-matrix method. 

The outline of our work is as follows: first, we will introduce the concept of HM
as well as the basic expressions needed to calculate the radiative heat flux for such materials 
in Sec.~II and III. In Sec.~IV we will discuss in detail the near-field heat transfer between mHMM. 
Finally in Sec.~V we summarize the main results and give some outlook to future work on this topic.

%
%
%

\section{Hyperbolic Media}

HM are a special class of anisotropic uni-axial media. In such anisotropic uni-axial media the permittivity tensor with respect to 
the principal axis can be written as a diagonal matrix of the form~\cite{Yeh}
\begin{equation}
  \uuline{\epsilon} = \begin{pmatrix} \epsilon_\perp & 0 & 0 \\ 0 & \epsilon_\perp & 0 \\ 0 & 0 & \epsilon_\parallel \end{pmatrix}.
\end{equation}
The permittivity $\epsilon_\parallel$ is the permittivity 'seen' by an electric field which is 
parallel to optical axis which coincides here with the z-axis, whereas the permittivity $\epsilon_\perp$
is the permittivity 'seen' by an electric field which is perpendicular to the optical axis. For the sake of
clarity let us in this section neglect dissipation, i.e.\ we assume that the permittivities are 
purely real numbers. Then one can easily classify different types of media~\cite{HuChui2002,Smith2003}:
\begin{enumerate}
  \item If $\epsilon_\perp > 0$ and $\epsilon_\parallel > 0$ then the medium is dielectric. For
        $\epsilon_\perp \neq \epsilon_\parallel$ the medium is uni-axial and for $\epsilon_\perp = \epsilon_\parallel$ it is isotropic.
  \item If $\epsilon_\perp < 0$ and $\epsilon_\parallel < 0$ then the medium is metallic. Again, for 
        $\epsilon_\perp \neq \epsilon_\parallel$ the medium is uni-axial and for $\epsilon_\perp = \epsilon_\parallel$ it is isotropic.
  \item If $\epsilon_\perp > 0$ and $\epsilon_\parallel < 0$ then the medium is called a type I hyperbolic material.
  \item If $\epsilon_\perp < 0$ and $\epsilon_\parallel > 0$ then the medium is called a type II hyperbolic material.
\end{enumerate}
That means that we can define HM by the property $\epsilon_\perp \epsilon_\parallel < 0$.
An extension towards uni-axial magnetic materials with hyperbolic properties is straightforward~\cite{HuChui2002,Smith2003}.

\begin{figure}
  \subfigure[type I]{\epsfig{file = 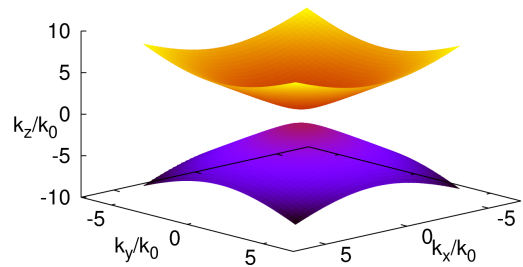, width = 0.45\textwidth}}
  \subfigure[type II]{\epsfig{file = 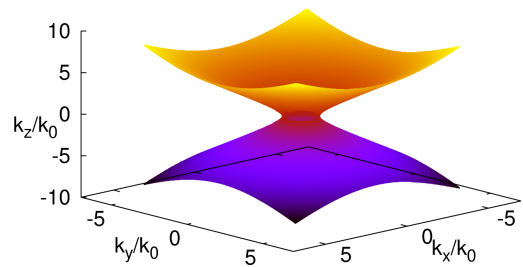, width = 0.45\textwidth}}
  \caption{Plot of the isofrequency curves for the extra-ordinary modes in k-space for 
           a (a) type I hyperbolic material with $\epsilon_\parallel = -0.7$ and $\epsilon_\perp = 1$
           and a (b) type II hyperbolic material with $\epsilon_\parallel = 0.7$ and $ \epsilon_\perp = -1$.\label{Fig:Hyperbolic}}
\end{figure}

In order to see the difference in the optical properties of the different classes of uni-axial materials,
one can solve for the eigensolutions of the vector-wave equation of the electromagnetic fields which will
give the dispersion relations of the two linear independent eigenmodes, the so-called ordinary and extra-ordinary
modes. The dispersion relation of the ordinary modes is given by~\cite{Yeh}
\begin{equation}
  \frac{k^2_x + k_y^2 + k_z^2}{\epsilon_\perp} - \frac{\omega^2}{c^2}  = 0
\label{Eq:DisperionOrdinary}
\end{equation}
and the dispersion relation of the extra-ordinary modes is given by~\cite{Yeh}
\begin{equation}
  \frac{k_x^2 + k_y^2}{\epsilon_\parallel} +  \frac{k_z^2}{\epsilon_\perp} - \frac{\omega^2}{c^2}  = 0.
\label{Eq:DisperionExtraOrdinary}
\end{equation}
From these relations it is obvious that only the extra-ordinary modes are subjected to the
anisotropy, since both permittivities $\epsilon_\perp$ and $\epsilon_\parallel$ enter in 
their dispersion relation. As a consequence only for these modes the difference between dielectric and
hyperbolic modes can be seen. This difference can be made obvious when fixing the frequency and plotting
the dispersion relations in k-space. These iso-frequency curves are for dielectrics ellipsoids. On the
other hand, for a type I (type II) hyperbolic material these iso-frequency curves are two-sheeted (one-sheeted)
hyperboloids as sketched in Fig.~\ref{Fig:Hyperbolic}. From these iso-frequency curves the main feature of 
HM becomes apparent: In principle, HM allow for propagating wave solutions
with arbitrarily large wave numbers, whereas for dielectric media the wave-vectors are bounded by
$\max(\sqrt{\epsilon_\parallel} k_0, \sqrt{\epsilon_\perp} k_0)$ where $k_0 = \omega/c$ is the modulus of the wave-vector
in vacuum. Of course, in practice also the allowed wave-vectors inside HM are limited by the properties of the material's crystalline
structure~\cite{YanEtAl2012,DrachevEtAl2013}, but they are generally speaking larger than in dielectric uni-axial media. Note that
there are no propagating wave solutions for the metallic case {because $\epsilon_\perp < 0$ and $\epsilon_\parallel < 0$. 
Since $\epsilon_\perp < 0$} for the type II HM, there are also no ordinary-mode solutions for type II HM.

One might think that materials fulfilling the condition $\epsilon_\perp \epsilon_\parallel < 0$ are
rare to find, but in fact this is not true. There are many different natural bulk HM
such as hexagonal boron nitride, tetradymites, etc.\ which have hyperbolic bands in the visible, near-infrared or inrared regime. For an overview
of different natural HM listing the different respective hyperbolic bands we refer the reader to 
the references~\cite{ThompsonEtAl1998,CaldwellEtAl2014,EsslingerEtAl2014,NarimanovKildishev2015,KorzebEtAl2015}. Apart from natural 
HM there also exists the possibility to artificially 
fabricate hyperbolic meta-materials using nanostructuration methods. Such hyperbolic meta-materials can be multilayer
structures consisting of alternating layers of dielectric and metallic slabs (mHMM) or of periodic metallic nanorod structures
which are immersed in a dielectric host (wHMM) with a subwavelength periodicity, for instance. The advantage of such
artificial structures is that the broadness and spectral position of the hyperbolic bands can be tailored 
to some extend by chosing different materials and different filling fractions of the combined materials. 

\begin{figure}
  \epsfig{file = 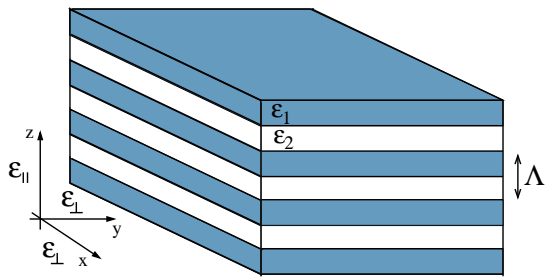, width = 0.4\textwidth}

  \caption{Sketch of a hyperbolic multilayered periodic structure with a period $\Lambda$ composed of materials with permittivities $\epsilon_1$ and $\epsilon_2$.
           \label{Fig:Multilayer}}
\end{figure}

In order to have a simple model for the permittivities of hyperbolic meta-materials one can use the
well-known expressions of effective medium theory (EMT). For example the effective permittivities of mHMM as
sketched in Fig.~\ref{Fig:Multilayer} can be modelled in the long-wavelength regime by~\cite{Yeh}
\begin{align}
  \epsilon_\perp &= f \epsilon_{\rm 1} + (1 - f)\epsilon_{\rm 2}, \label{Eq:epsperp} \\
  \epsilon_\parallel &= \biggl( \frac{f}{\epsilon_{\rm 1}} + \frac{1 - f}{\epsilon_{\rm 2}} \biggr)^{-1}, \label{Eq:epsparallel}
\end{align}
where $f$ is the volume filling fraction of material $1$ with permittivity $\epsilon_1$ and $1 - f$ is the volume
filling fraction of material $2$ with permittivity $\epsilon_2$. These expressions, {which can be obtained
by an averaging of the electric field over one period of the structure,} are strictly speaking only meaningful when
the periodicity $\Lambda$ of the meta-material structure is much smaller than the wavelength but even for $\Lambda$ much smaller than
the wavelength the EMT results can lead to wrong predictions~\cite{OrlovEtAl2011,KidwaiEtAl2011,KidwaiEtAl2012,IorshEtAl2012,ChebykinEtAl2012,TschikinEtAl2013,Biehs2} so that some care needs to be exercised when using these EMT expressions. Nonetheless, they are quite helpful for determinig the
hyperbolic frequency bands of a meta-material structure in a simple way. 

\begin{figure}
  \subfigure{\epsfig{file = 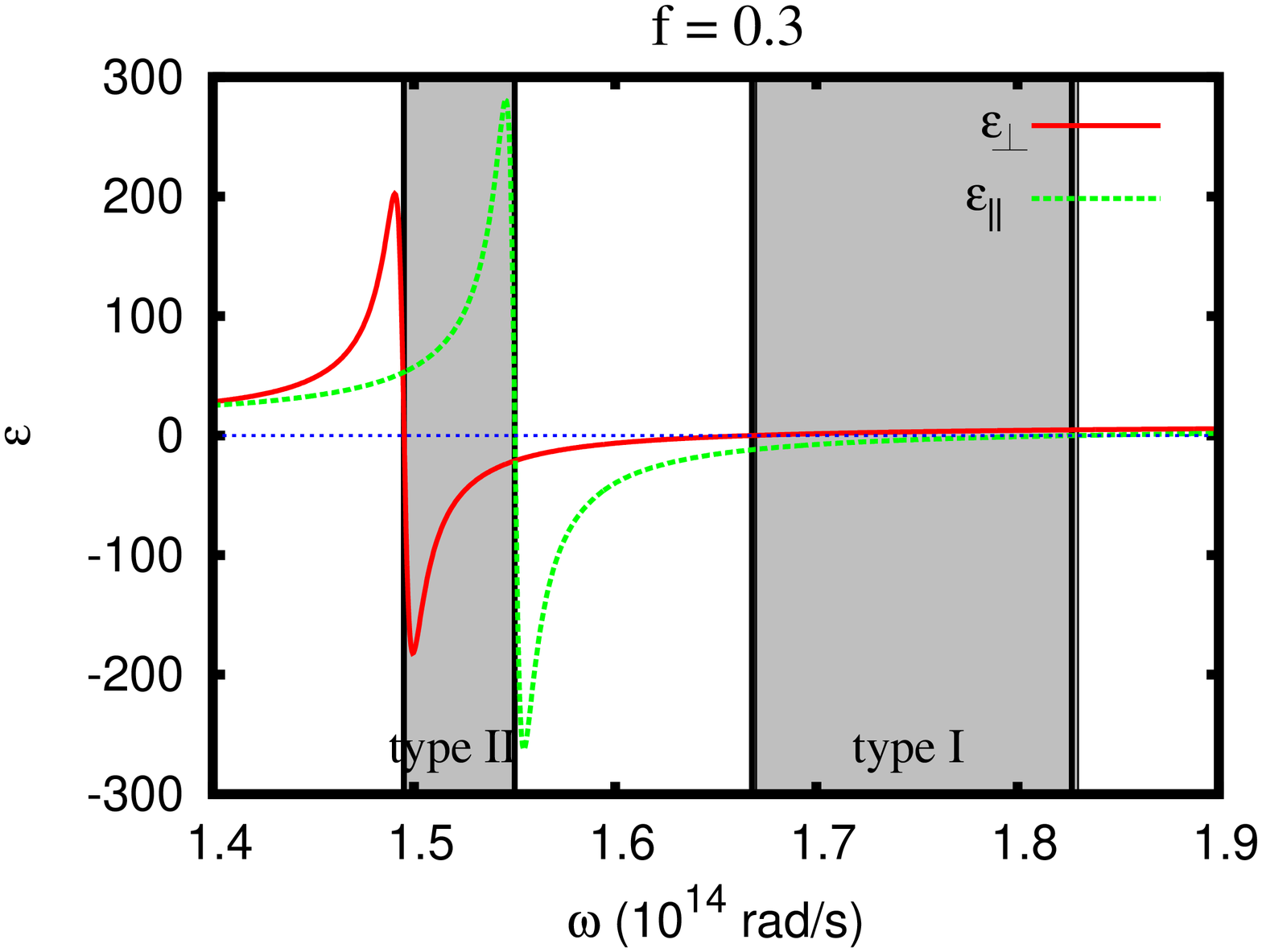, width = 0.4\textwidth}}
  \subfigure{\epsfig{file = 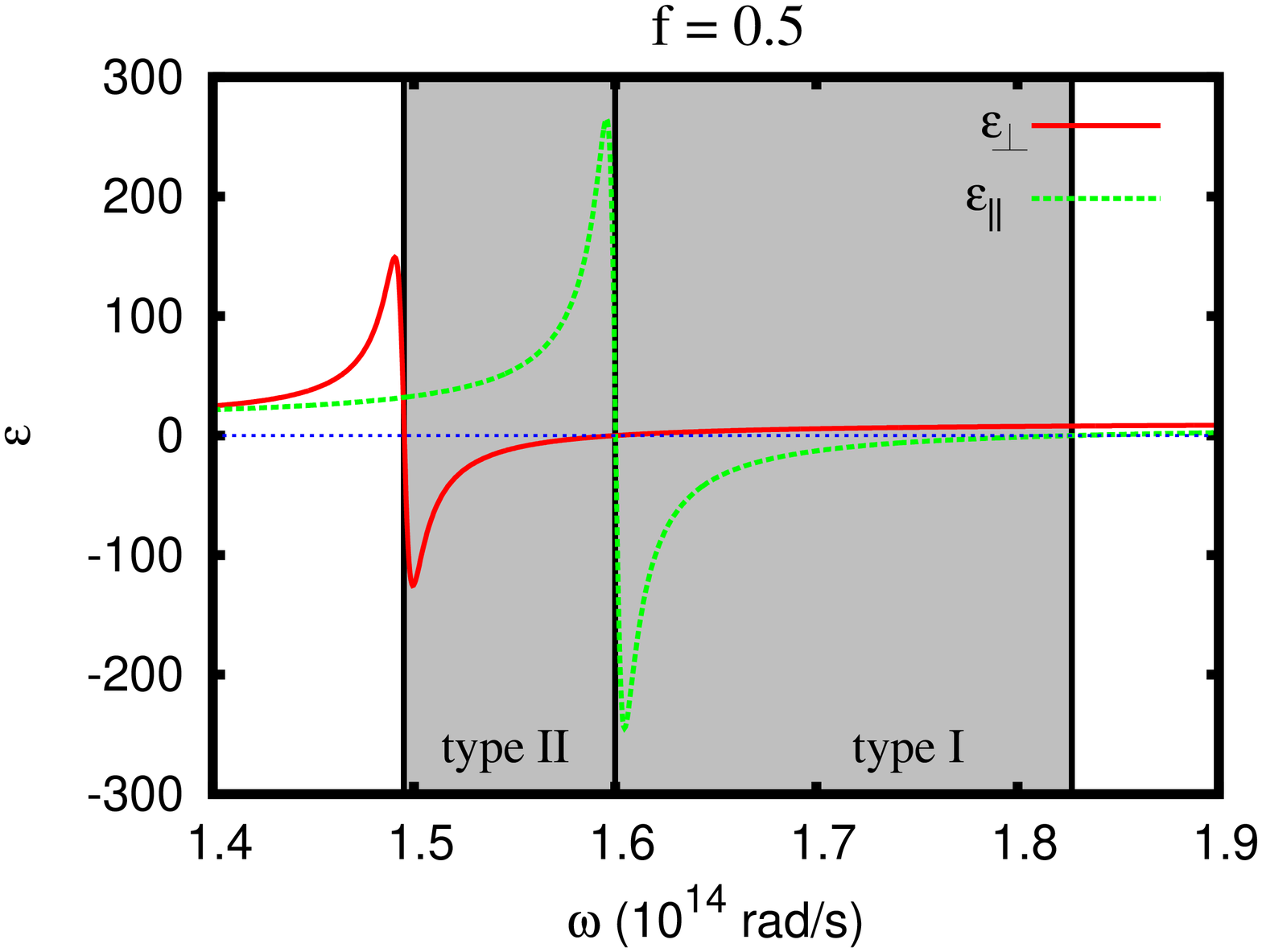, width = 0.4\textwidth}}
  \subfigure{\epsfig{file = 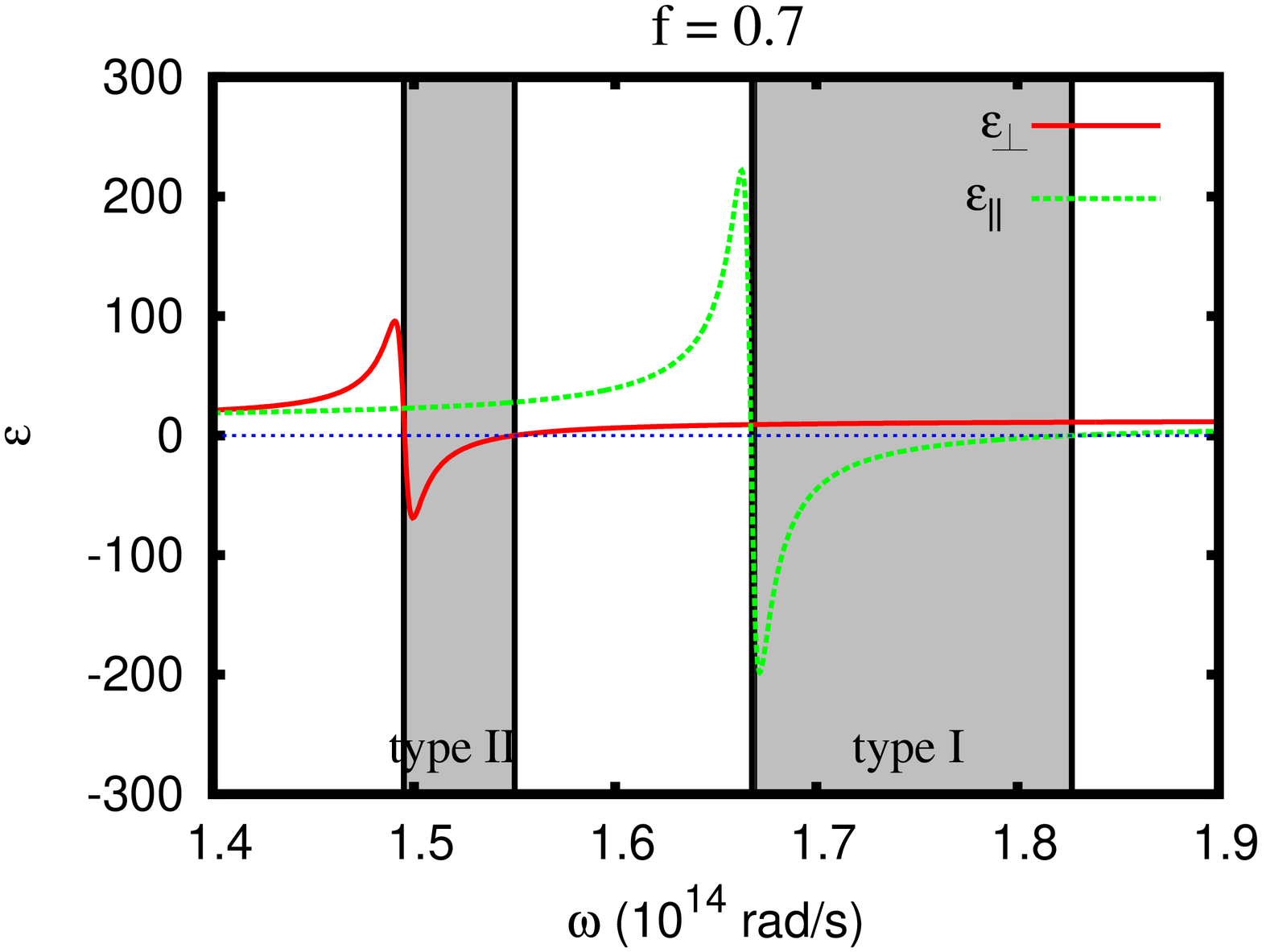, width = 0.4\textwidth}}

  \caption{Plot of the {real parts of the} permittivities $\epsilon_\parallel$ and $\epsilon_\perp$ for a SiC/Ge mHMM using the EMT
           expressions from Eqs.~(\ref{Eq:epsperp}) and (\ref{Eq:epsparallel}) for filling fractions $f = 0.3, 0.5, 0.7$ of SiC.
           The hyperbolic frequency bands of type I and II (in grey) lie in the reststrahlen band $\omega_{\rm TO} < \omega < \omega_{\rm LO}$
           of SiC where $\omega_{\rm TO} = 1.495\times10^{14}\,{\rm rad/s}$ and $\omega_{\rm LO} = 1.827\times10^{14}\,{\rm rad/s} $ are the frequencies of the transversal and longitudinal phonons in SiC.\label{Fig:Permittivity}}
\end{figure}

As an example, we show in Fig.~\ref{Fig:Permittivity} a plot of the real parts of the effective permittivites of a  SiC/Ge mHMM structure 
with $\epsilon_1 = \epsilon_{\rm SiC}$ and $\epsilon_2 = \epsilon_{\rm Ge} = 16$, where~\cite{Palik98}
\begin{equation}
   \epsilon_{\rm SiC} (\omega) =\epsilon_{\infty}\frac{\omega ^2-\omega_\text{LO}^2 + \ri\gamma\omega}{\omega^2-\omega_\text{TO}^2+ \ri\gamma\omega},
\end{equation}
with $\epsilon_{\infty}=6.7$, $\omega_{\rm TO}=1.495\cdot 10^{14}\,$rad/s, $\omega_{\rm LO}=1.827\cdot 10^{14}\,$rad/s and $\gamma=0.9\cdot 10^{12}\,$rad/s.
It can be seen that by changing the filling fraction $f$ for SiC the broadness of the hyperbolic 
bands can be controlled where the largest hyperbolic bands are obtained for $f = 0.5$. In this case both hyperbolic bands
are extended over the whole reststrahlen band of SiC~\cite{Bohren} in which SiC has metal-like properties, i.e.\ its permittivity
is negative. Of course, the overall hyperbolic frequency band can be made larger by using materials with a broader reststrahlen
band or by using two phonon-polaritonic materials like SiC and GaN with adjacent reststrahlen bands. Note that although metals
would have a negative permittivity over the whole frequency range below the plasma frequency of the metal, they are 
less advantageous than phonon-polaritonic materials mHMM structures in the infrared, because the losses become quite 
large for thin metallic slabs~\cite{LangEtAl2013}.

%
%
%

\section{Heat Flux Formula for anisotropic media}

\begin{figure}
  \epsfig{file = 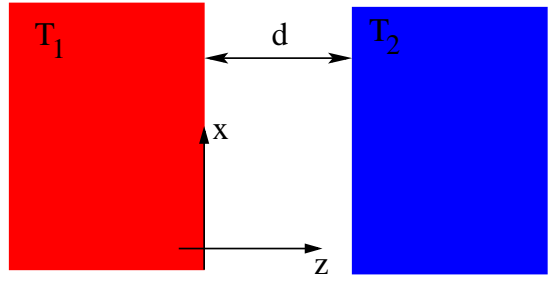, width = 0.4\textwidth}
  \caption{Sketch of two semi-infinite materials at two different temperatures $T_1$ and $T_2$ which are
           separated by a vacuum gap of thickness $d$.\label{Fig:SketchGeometry}}
\end{figure}

In order to determine the heat flux between two HM at temperatures $T_1$ and $T_2$ which are separated by vacuum
gap of thickness $d$ as sketched in Fig.~\ref{Fig:SketchGeometry} the heat flux expression of Polder and van Hove~\cite{PvH1971} which is based on Rytov's 
fluctuational electrodynamics has been generalized to anisotropic media~\cite{Bimonte,Opt_Exp,Messina2011,Krueger2012}. The resulting heat flux 
expression can be written as
\begin{equation}
  \Phi =  \int_0^\infty\!\frac{\rd \omega}{2 \pi} \bigl[ \Theta(\omega,T_1) - \Theta(\omega,T_2)\bigr] \Phi_\omega.
\end{equation}
where ($i = 1,2$)
\begin{equation}
  \Theta(\omega,T_i) = \frac{\hbar \omega}{\re^{\frac{\hbar \omega}{\kb T_i}} - 1} ,
\end{equation}
is the thermal part of the  mean energy of a quantum mechanical harmonic oscillator at temperature $T_i$;
$2 \pi \hbar$ is Planck's constant and $\kb$ is Boltzmann's constant. For the {reduced} spectral heat flux we obtain
\begin{equation}
  \Phi_\omega = \int\!\!\frac{\rd^2\kappa}{(2 \pi)^2} \, \mathcal{T}(\omega,\boldsymbol{\kappa}; d)
\label{Eq:SpectralPoynting}
\end{equation}
introducing the transmission coefficient
\begin{equation}
   \mathcal{T}(\omega,\boldsymbol{\kappa}; d) =  
    \begin{cases}
      \scriptstyle\Tr\bigl[(\mathds{1} - \mathds{R}_2^\dagger \mathds{R}_2)  \mathds{D}^{12}(\mathds{1} - \mathds{R}_1 \mathds{R}_1^\dagger)  {\mathds{D}^{12}}^\dagger \bigr], &   \!\!\!\!\!\! \scriptstyle \kappa < k_0\\
     \scriptstyle \Tr\bigl[(\mathds{R}_2^\dagger - \mathds{R}_2) \mathds{D}^{12} (\mathds{R}_1 - \mathds{R}_1^\dagger)  {\mathds{D}^{12}}^\dagger \bigr]\re^{-2 |\gamma_{0}| d} ,  & \!\!\!\!\!\! \scriptstyle \kappa > k_0
  \end{cases}
\end{equation}
which has the property $\mathcal{T} \in [0,2]$ {which is due to the fact, that the so defined transmission coefficients accounts for both polarizations as well as for possible depolarization effects.} The 2x2 reflection matrix of the two interfaces are given by ($i = 1,2$)
\begin{equation}
\label{ReflectionMatrices}
{\mathds R}_i = \begin{pmatrix}
   r^{{\rm ss}}_i (\omega, \kappa) &  r^{{\rm sp}}_i (\omega, \kappa) \\
   r^{{\rm ps}}_i (\omega, \kappa) &  r^{{\rm pp}}_i (\omega, \kappa) 
  \end{pmatrix}.
\end{equation}
The Fresnel reflection coefficients $r^{{\rm ss}}_i (\omega, \kappa)$, $ r^{{\rm ps}}_i (\omega, \kappa)$ etc.\ 
are the usual reflection coefficients of interface $i$ describing the scattering of an incoming s-polarized plane 
wave into a s-polarized or p-polarized wave etc. The Fabry-P\'{e}rot-like 'denominator' is defined as
\begin{equation}
  \mathds{D}^{12} = (\mathds{1} - \mathds{R}_1 \mathds{R}_2 \re^{2 \ri \gamma_{0} d})^{-1}.
\end{equation}
Here $\gamma_0 = \sqrt{\kappa^2 - k_0^2}$ is the wave-vector component normal to the interfaces of the two anisotropic materials
and $\boldsymbol{\kappa} = (k_x,k_y)^t$ is the wave-vector component parallel to the interfaces (see Fig.~\ref{Fig:SketchGeometry}).
For more details on the derivation and definition of the different quantities we refer the reader to Ref.~\cite{Opt_Exp}, for instance.
From the above expressions it becomes clear that the transmission coefficient $\mathcal{T}$ takes into account
the propagating wave contributions ($\kappa < k_0$) as well as the evanescent wave contributions ($\kappa > k_0$). 
The evanescent wave contribution is the reason for having super-Planckian heat transfer at the nanoscale~\cite{PvH1971}. On the other hand,
for the propagating wave contribution the blackbody limit cannot be overcome as has been proved once more for anisotropic media
in the framework of fluctuational electrodynamics in Ref.~\cite{Biehs2016}, recently. 

Finally, in the special case where we consider two uni-axial materials having an optical axis which is normal to the
interface the above expressions simplify, because in this case the reflection matrix becomes diagonal, i.e.\
we have $r^{\rs \rp}_i = r^{\rp \rs}_i = 0$, with the diagonal components
\begin{align}
    r^{\rs \rs}_i &=  \frac{\gamma_{0} - \gamma_{i,\rs}}{\gamma_{0} + \gamma_{i,\rs}} \equiv r^{{\rs}}_i (\omega, \kappa),  \\
    r^{\rp \rp}_i &=  \frac{\epsilon_{i,\parallel} \gamma_{0} - \gamma_{i,\rp}}{\epsilon_{i,\parallel} \gamma_{0} + \gamma_{i,\rp}} \equiv r^{{\rp}}_i (\omega, \kappa).  
\end{align}
Furthermore, when the optical axis coincides with the surface normal the ordinary (extra-ordinary) modes coincide
with the s-polarized (p-polarized) modes. Hence,  the z-components of the wave-vector for s- and p-polarization 
$\gamma_{i,\rs/\rp}$ are determined by the dispersion relations in Eqs.~(\ref{Eq:DisperionOrdinary}) and (\ref{Eq:DisperionExtraOrdinary})
yielding
\begin{align}
  \gamma_{i,\rs} &= \sqrt{\epsilon_{i, \perp} k_0^2 - \kappa^2} ,\\
  \gamma_{i,\rp} &= \sqrt{\epsilon_{i, \perp} k_0^2 - \frac{\epsilon_{i,\perp}}{\epsilon_{i,\parallel}}\kappa^2}.
\end{align}
Then the transmission coefficient simplifies to $\mathcal{T} = \mathcal{T}_\rs + \mathcal{T}_\rp$ where
the transmission coefficients for the two polarization states of light are given by ($\lambda = \rs,\rp$)
\begin{equation} 
  \mathcal{T}_\lambda (\omega,\boldsymbol{\kappa}; d) = 
     \begin{cases}
                \frac{(1 - |r^{\lambda}_1|^2)(1 - |r^{\lambda}_2|^2)}{|1 - r^{\lambda}_1 r^{\lambda}_2 \exp(2 {\rm i} \gamma_0 d)|^2}, & \!\!\!\!\! \kappa < k_0 \\ \vspace{0.2cm}
                4\frac{ \Im(r^{\lambda}_1) \Im(r^{\lambda}_2) e^{-2 |\gamma_0| d} }{|1 - r^{\lambda}_1 r^{\lambda}_2 \exp(2 {\rm i} \gamma_0 d)|^2},  & \!\!\!\!\!\kappa > k_0  
     \end{cases}
  \label{Eq:TransmissionCoeff}
\end{equation}
with $ \mathcal{T}_\lambda \in [0,1]$. By setting $\epsilon_{i,\perp} = \epsilon_{i,\parallel} \equiv \epsilon_i$ we retrieve the well-known results 
of Polder-van Hove~\cite{PvH1971} describing the heat flux between two isotropic semi-infinite materials with permittivities $\epsilon_i$.
The same expressions can of course also be used for magneto-optical uni-axial materials by adding the magnetic properties to the reflection coefficients~\cite{WuEtAl2015,ChangEtAl2016}.

%
%
%
%
%

\begin{figure*}
  \subfigure[N = 0]{\epsfig{file = 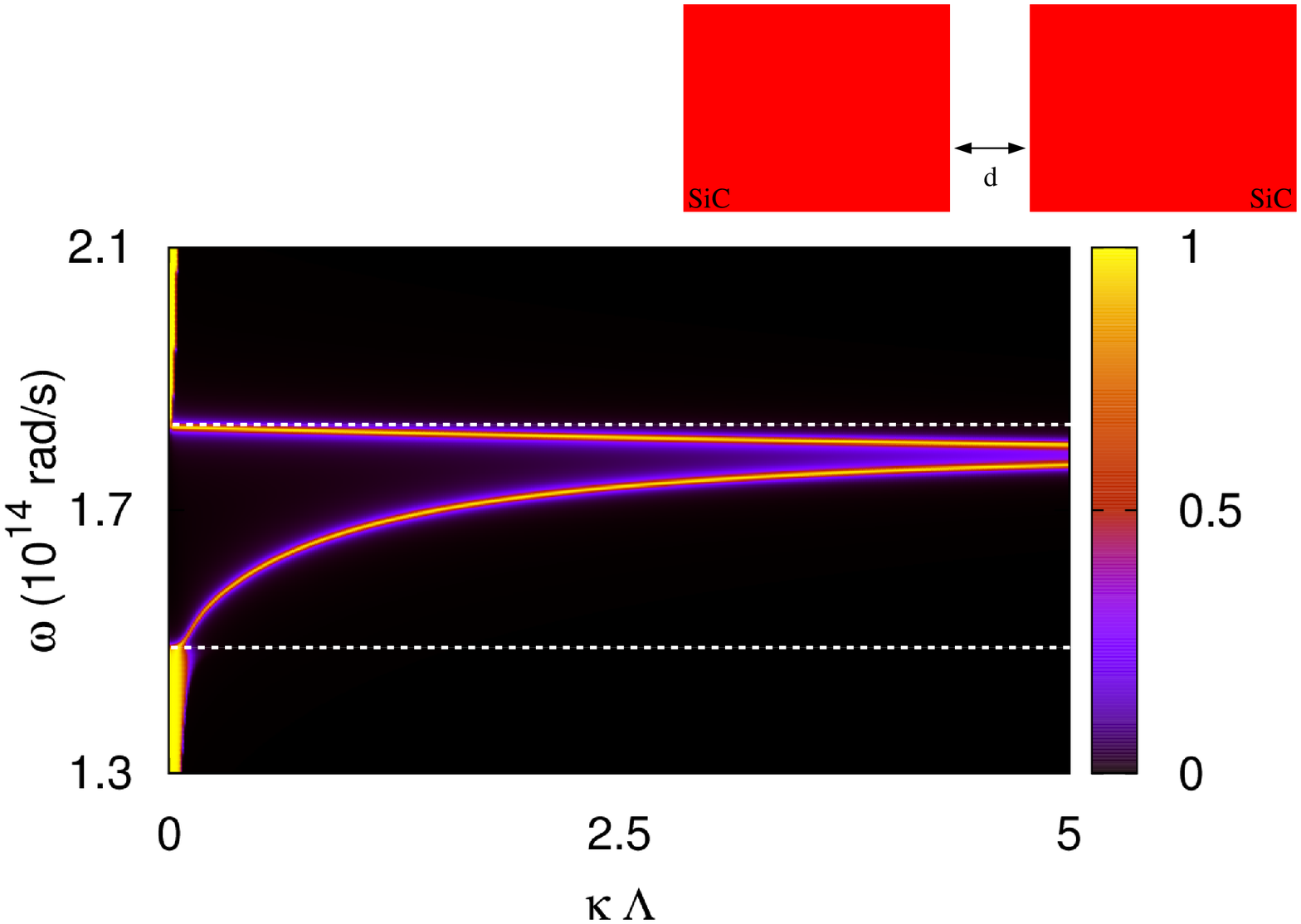, width = 0.3\textwidth}}
  \subfigure[N = 1] {\epsfig{file = 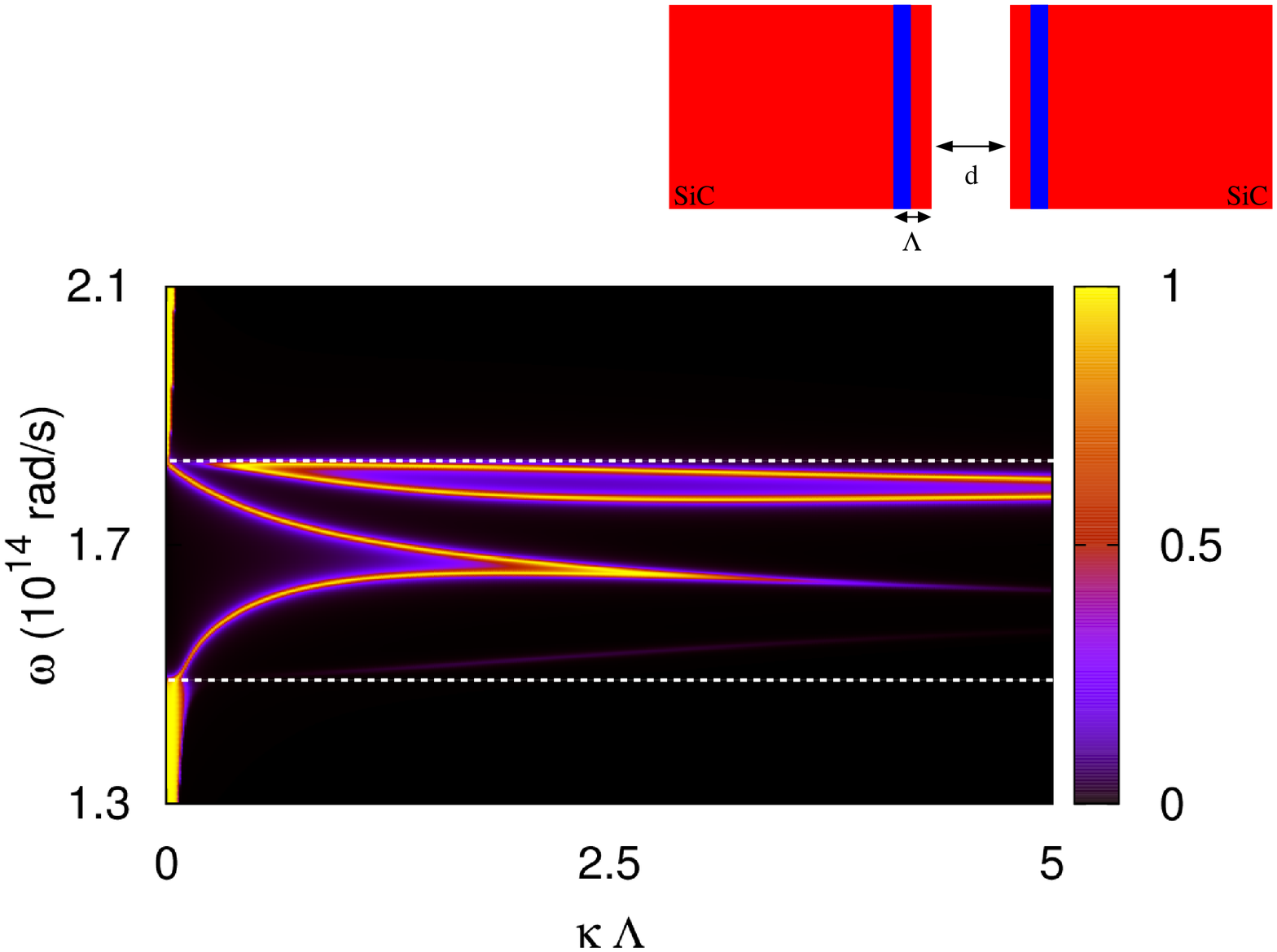, width = 0.3\textwidth}}
  \subfigure[N = 2]{\epsfig{file = 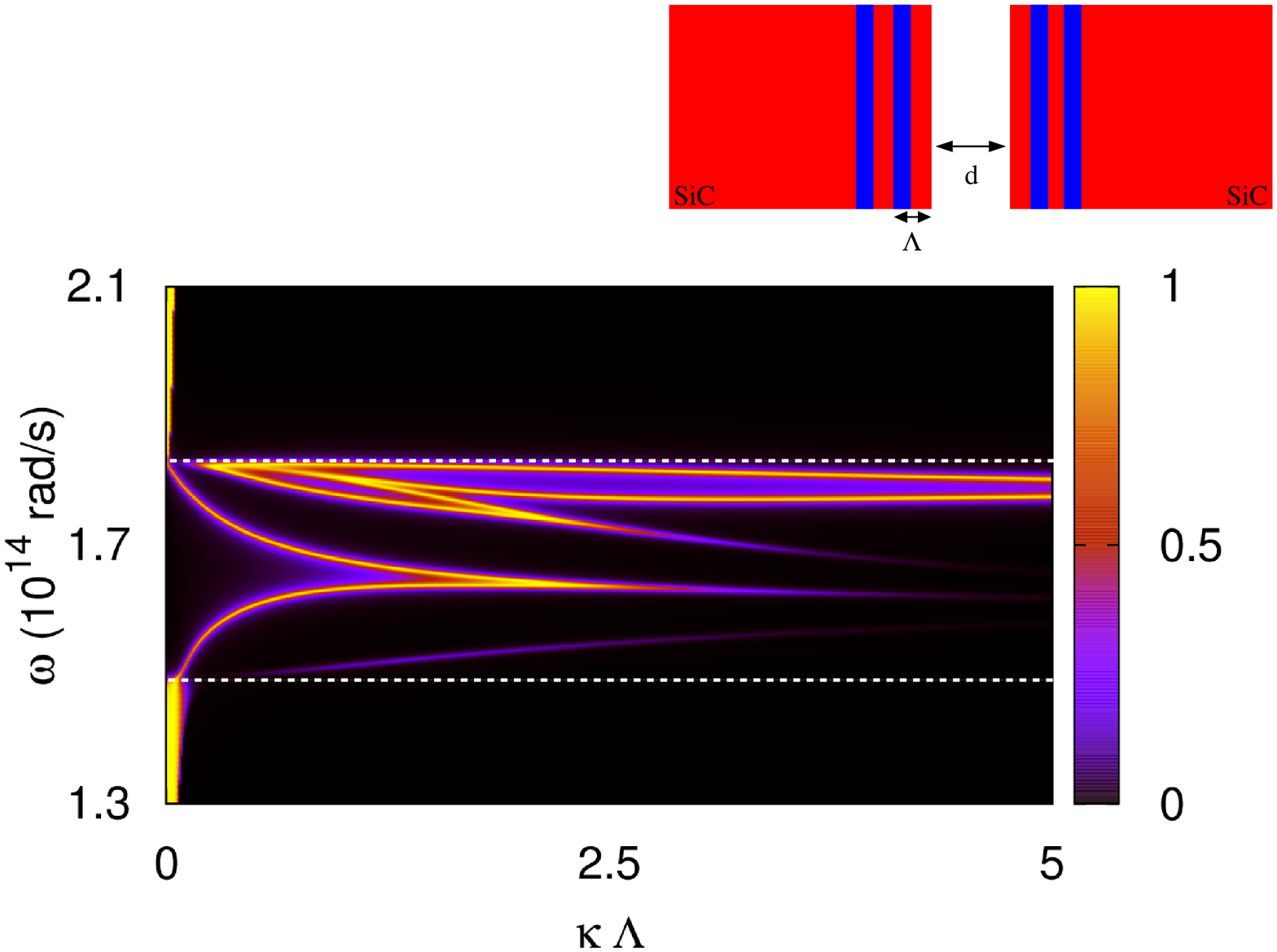, width = 0.3\textwidth}}
  \subfigure[N = 3]{\epsfig{file = 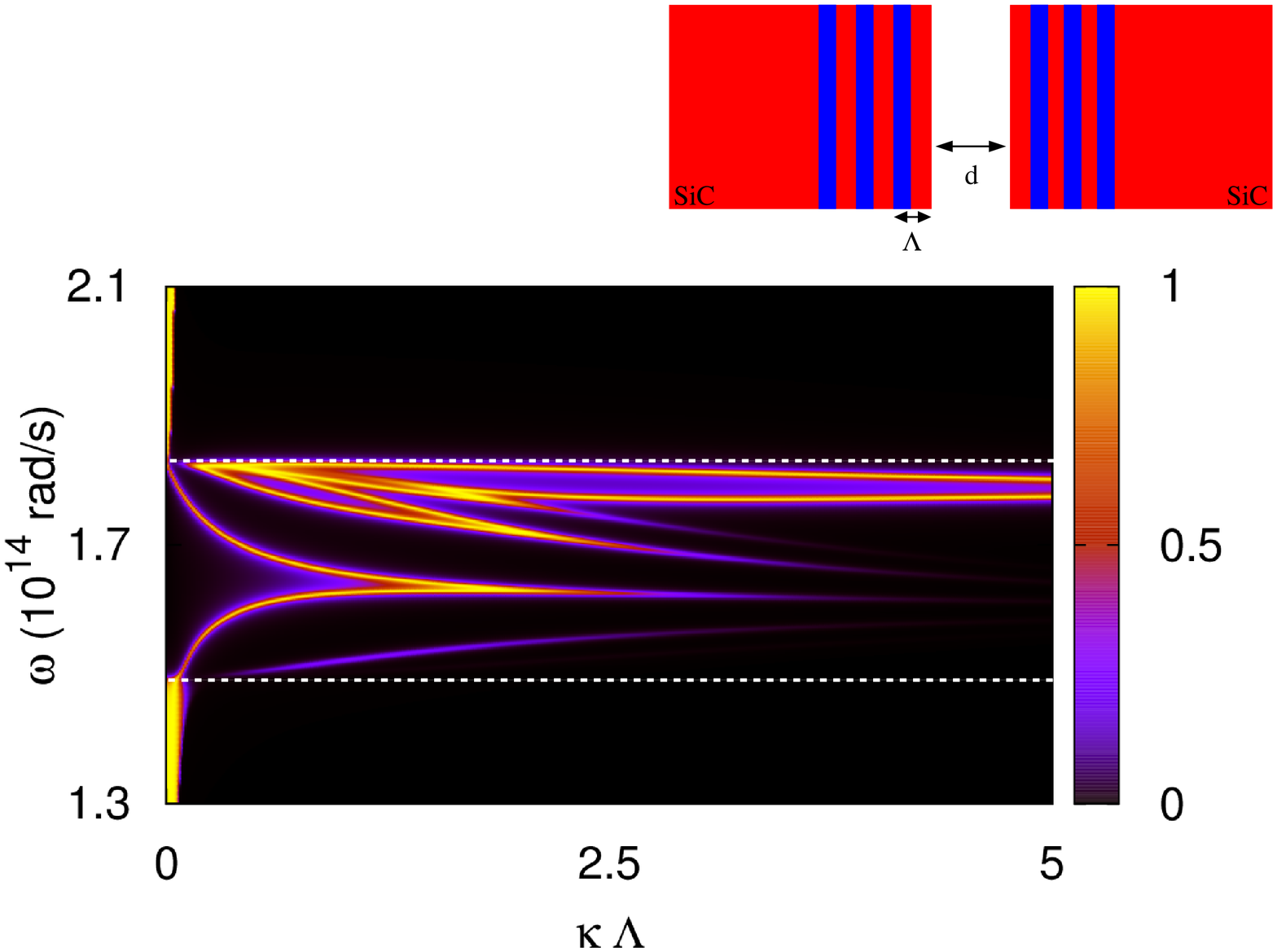, width = 0.3\textwidth}}
  \subfigure[N = 4]{\epsfig{file = 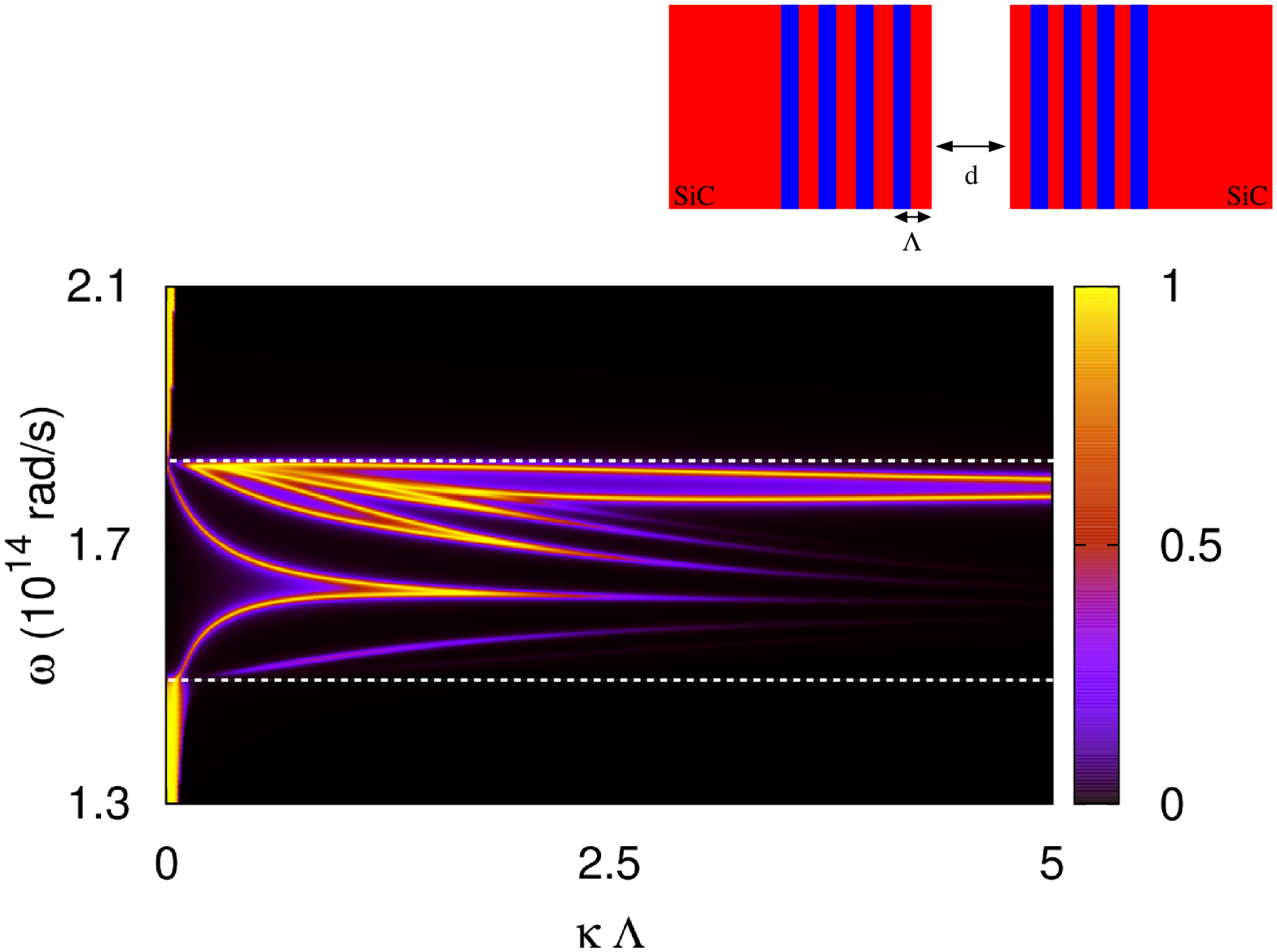, width = 0.3\textwidth}}
  \subfigure[N = 9]{\epsfig{file = 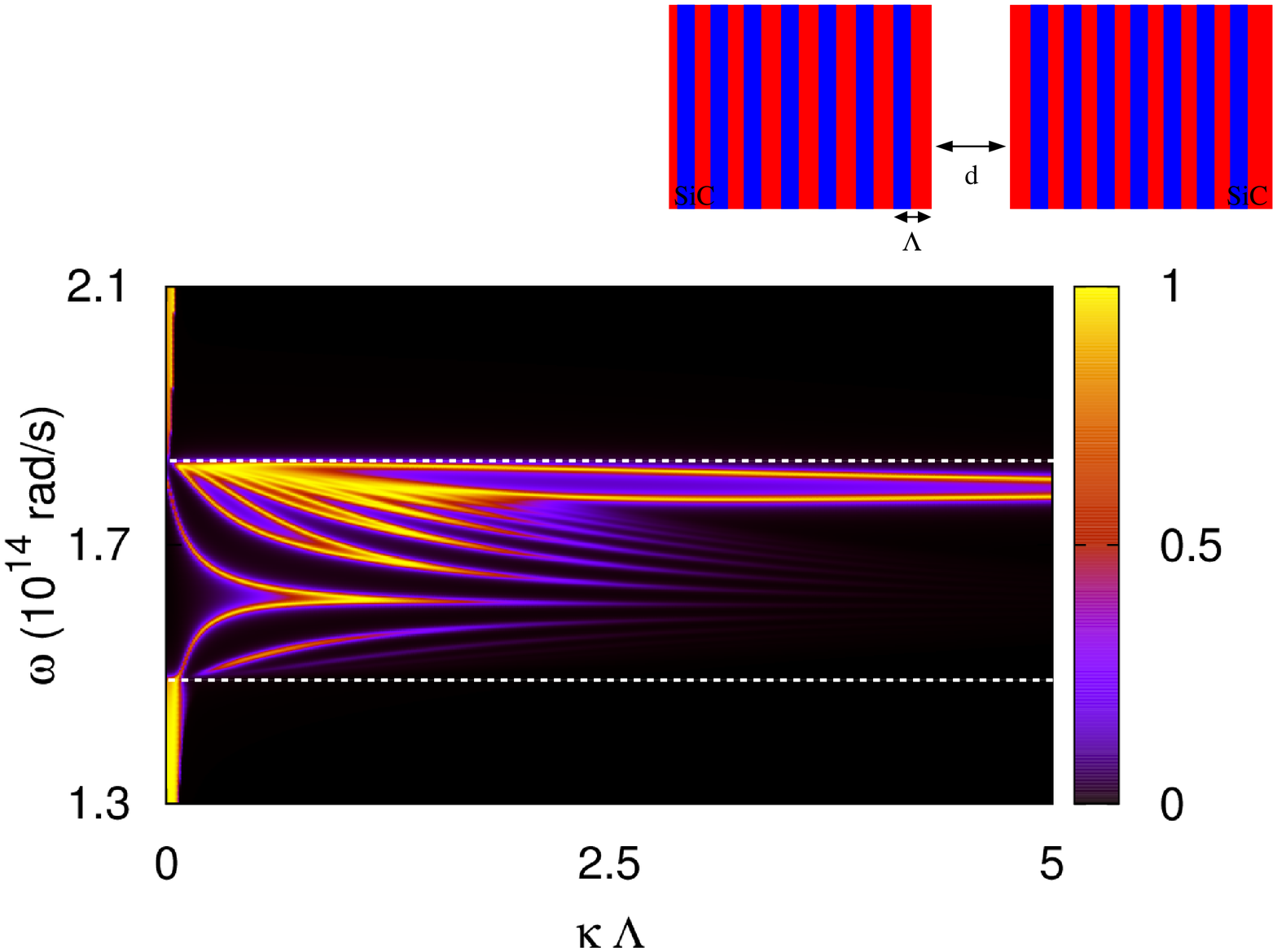, width = 0.3\textwidth}}
  \caption{Transmission coefficient $\mathcal{T}_\rp$ of the p-polarized modes in Eq.~(\ref{Eq:TransmissionCoeff}) in $(\omega,\kappa)$ space
           fixing $d = 10\,{\rm nm}$. The horizontal lines represent $\omega_{\rm TO}$ and $\omega_{\rm LO}$ of SiC, i.e.\ in between
           both lines lies the reststrahlenband.
           (a) two SiC bulk media. (b) first unit cell of the mHMM is achieved by adding a $10\,{\rm nm}$ thick germanium layer 10nm below the SiC
           surface. This Ge/SiC layer of thickness $\Lambda = 20\,{\rm nm}$ is the first unit-cell of an infinite Ge/SiC hMM with 
           period $\Lambda$ and filling fraction $f = 0.5$. (c)-(f) further germanium layers of $10\,{\rm nm}$ thickness are 
           added until we have a mHMM of 9 SiC/Ge unit cells or 18 layers with an overall thickness of 180nm. Here $N$ stands for the number
           of SiC/Ge unit cells.~\label{Fig:TransmissionCoefficient}}
\end{figure*}

\section{Multilayer HMM}

\subsection{Formation of hyperbolic bands}

First, let us see how the hyperbolic bands are formed in mHMM. To this end, we first chose a
 SiC/Ge mHMM with a filling fraction of $f = 0.5$, because in this case the hyperbolic bands
are the broadest as shown in Fig.~\ref{Fig:Permittivity}. Furthermore, we choose 
$10\,{\rm nm}$ for the germanium and the SiC layers so that $\Lambda = 20\,{\rm nm}$ and we chose 
a vacuum gap of $d = 10\,{\rm nm}$. In order to show the formation of the hyperbolic bands we plot the 
transmission coefficient for p-polarized light in Eq.~(\ref{Eq:TransmissionCoeff}) for two SiC bulk materials first in Fig.~\ref{Fig:TransmissionCoefficient}(a). For SiC we choose $\epsilon_\parallel = \epsilon_\perp = \epsilon_{\rm SiC}$. Then we add successively $10\,{\rm nm}$ layers of Ge such that we obtain a mHMM with $N = 9$ SiC/Ge unit cells and $f = 0.5$. The reflection coefficients $r^\rs_i$ and $r^\rp_i$ for these structures
are calculated by the S-matrix method~\cite{AuslenderHava1996} and then used in Eq.~(\ref{Eq:TransmissionCoeff}) to calculate the
transmission coefficient $\mathcal{T}_\rp$. 

%
%

\begin{figure*}
  \subfigure[SiC on top]{\epsfig{file = 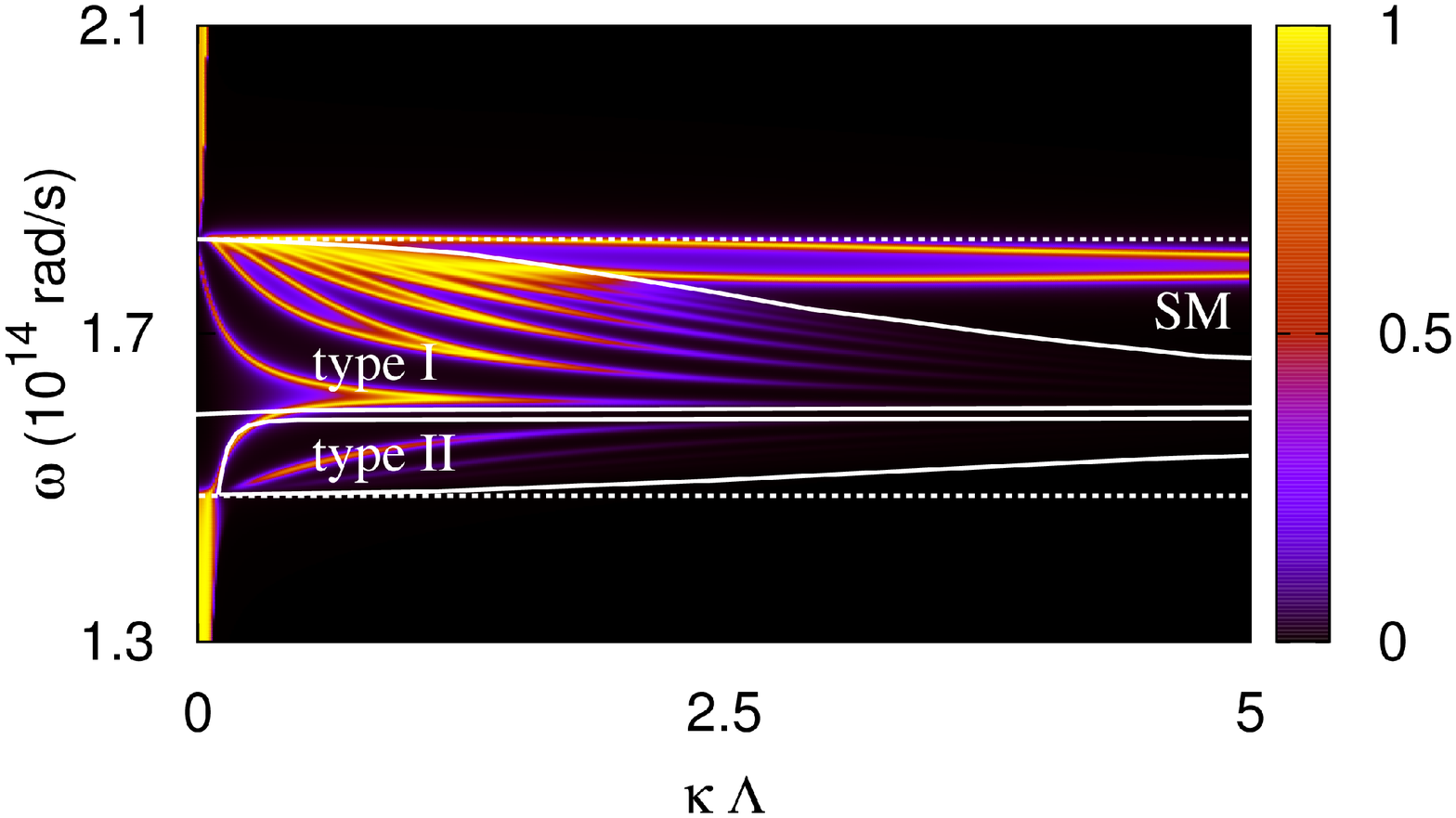, width = 0.45\textwidth}}
  \subfigure[Ge on top]{\epsfig{file = 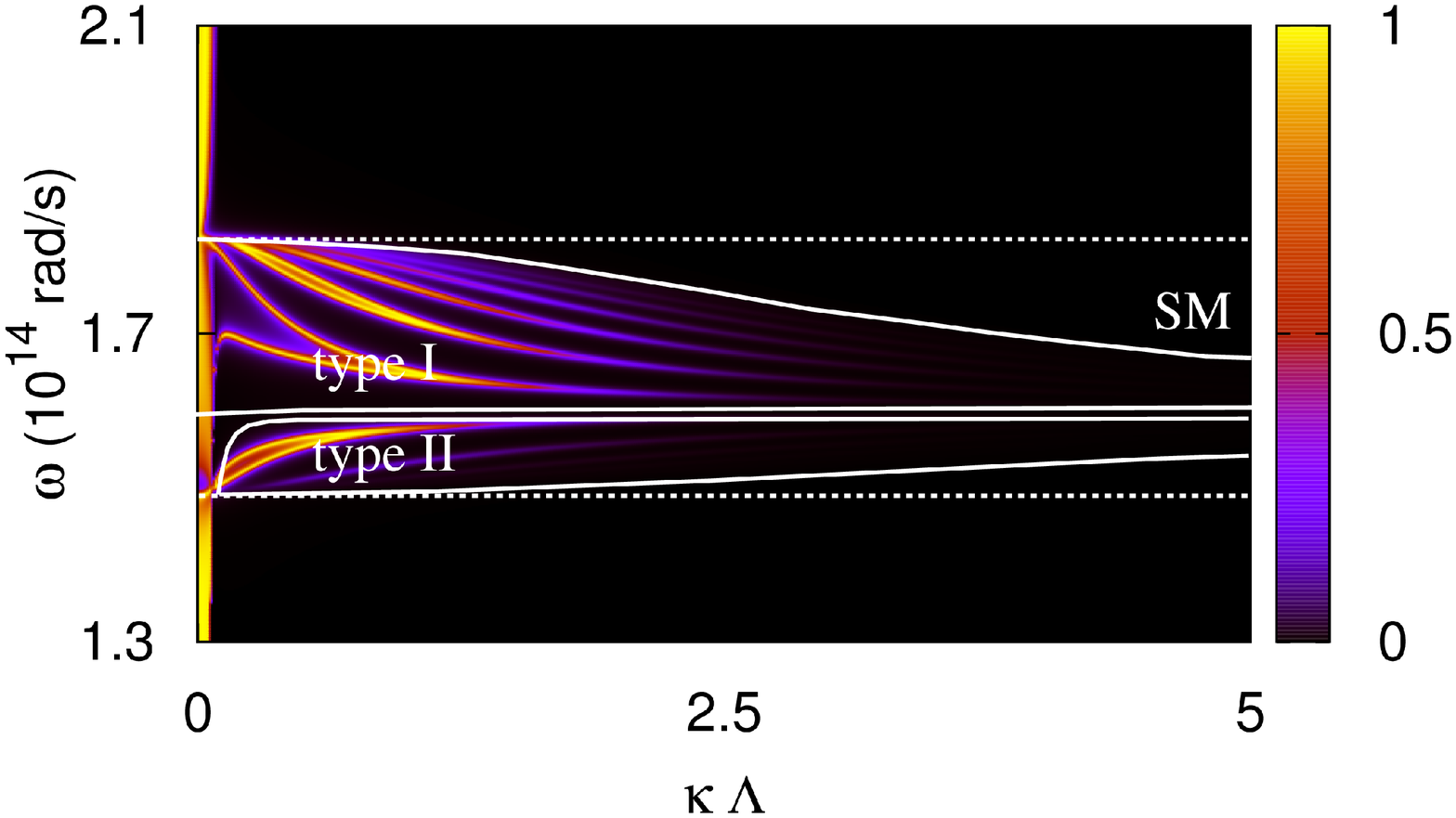, width = 0.45\textwidth}}
  \subfigure[mixed]{\epsfig{file = 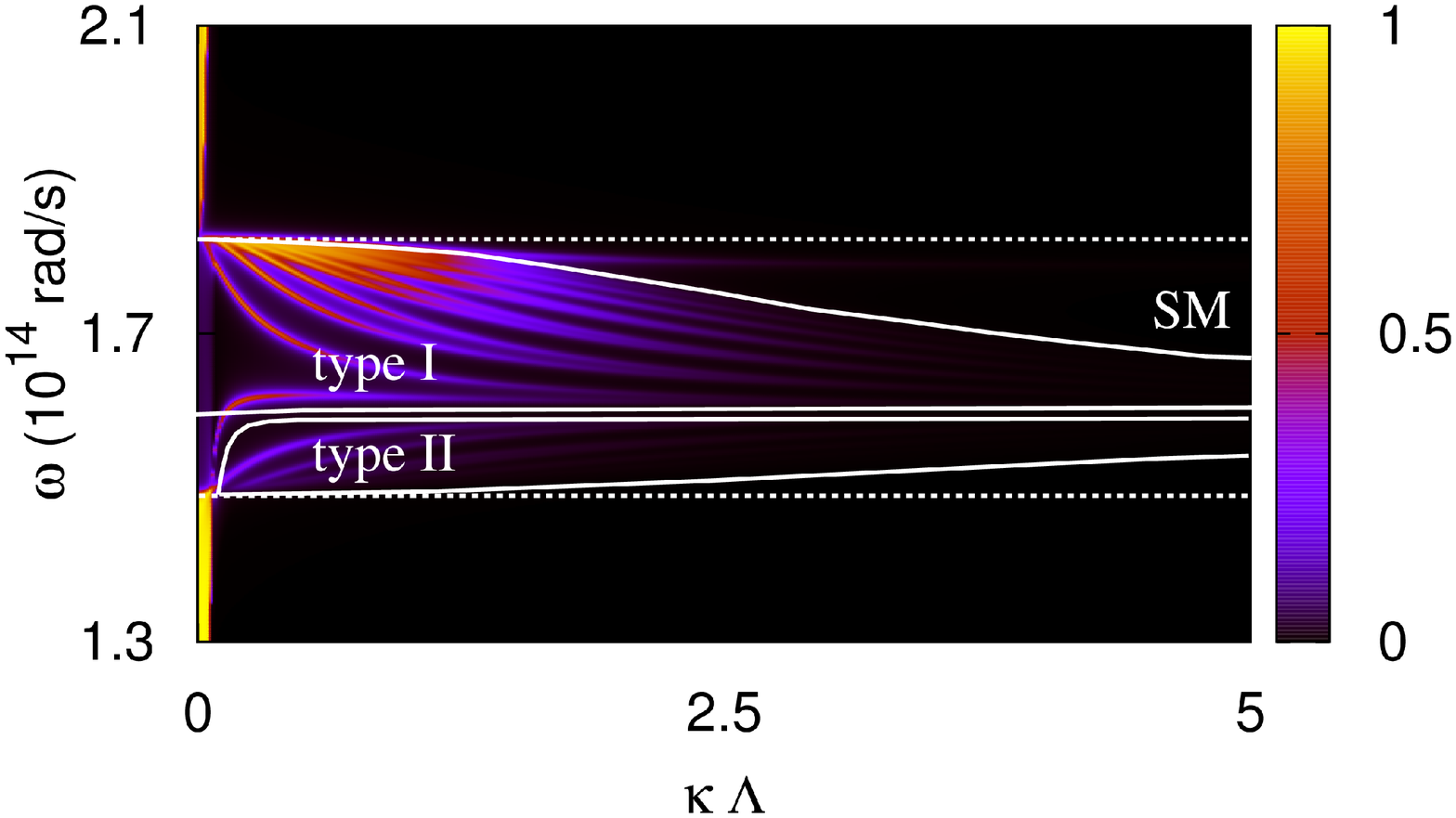, width = 0.45\textwidth}}
  \subfigure[EMT]{\epsfig{file = 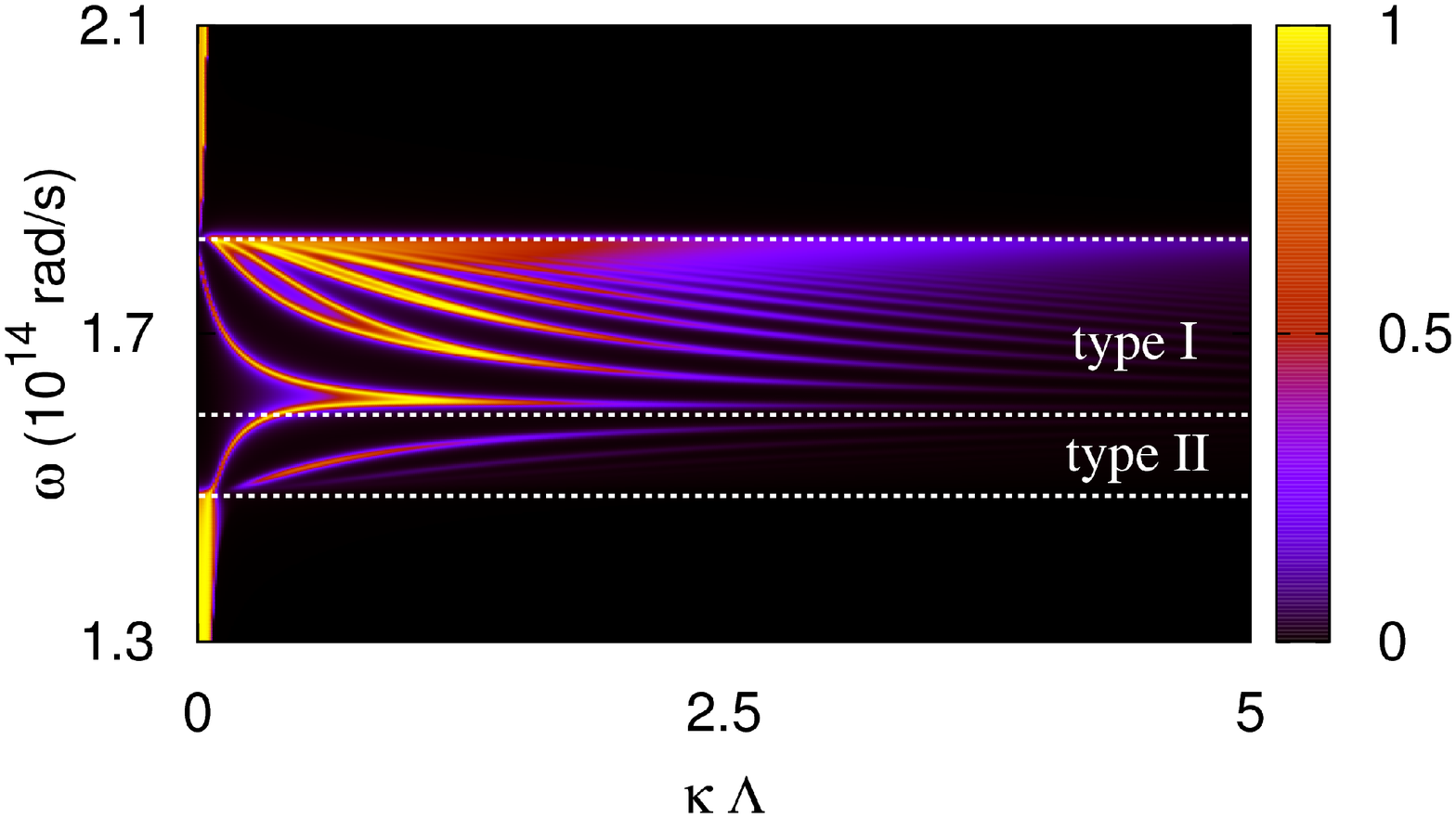, width = 0.45\textwidth}}
  \caption{Transmission coefficient $\mathcal{T}_\rp$ as in Fig.~\ref{Fig:TransmissionCoefficient} but for different configurations. The solid white lines
           mark the edges of the hyperbolic bands determined with the exact dispersion relation (\ref{Eq:BlochDisp}) for an infinite hyperbolic structure. Here
           the exact results are for a structure with $N = 9$ SiC/Ge unit cells, i.e.\ 18 layers, so that the hyperbolic bands are not completely filled but the
           different modes can be seen. \label{Fig:TransmissionCoefficient2}}
\end{figure*}

In Fig.~\ref{Fig:TransmissionCoefficient}(a) the surface modes of the two bulk SiC materials can be nicely seen. When
adding germanium layers in Fig.~\ref{Fig:TransmissionCoefficient}(b)-(f) we add successively new coupled modes in the spectral 
frequency window between $\omega_{\rm TO}$ and $\omega_{\rm LO}$ of SiC, i.e.\ the reststrahlenband. These 
coupled modes are the coupled surface phonon polaritons of the SiC layers which couple via the Ge layers~\cite{ZhukovskyEtAl2014}. 
They form new bands of modes in the reststrahlen band which are the hyperbolic bands. These hyperbolic bands
are the type I and type II bands. It is interesting to note that the group velocity in the direction parallel to
the interface of the modes in the type I band can be negative. This leads to negative or anomalous refraction~\cite{HuChui2002,Smith2003}
which has already been measured for mHMM~\cite{HoffmanEtAl2007}. This negative refraction does of course also occur 
for near-field thermal radiation and has been studied for instance in Ref.~\cite{BrightEtAl2014} by using the
energy-streamline method. 

From the shown plots in Fig.~\ref{Fig:TransmissionCoefficient} it is also apparent that the coupled surface mode resonances of 
the two bulk SiC materials which are coupled via the vacuum gap in Fig.~\ref{Fig:TransmissionCoefficient}~(a) persist when adding 
the layers of germanium. These surface modes can be formally distinguished from the hyperbolic modes: This 
is so because the hyperbolic modes are propagating inside the multilayer medium, i.e.\ they are Bloch modes, 
whereas the surface modes are evanescent in nature. For an infinite multilayer strucure the Bloch modes are 
fulfilling the dispersion relation~\cite{Yeh} ($\lambda = \rs, \rp$)
\begin{equation}
\begin{split}
  \cos(\gamma_\lambda \Lambda) &= - \frac{1}{2}\biggl(\frac{k_{z1}}{P_\lambda k_{z2}} + \frac{P_\lambda k_{z2}}{k_{z1}} \biggr) \sin(k_{z1} l_1) \sin(k_{z2} l_2)\\
                  &\quad + \cos(k_{z1} l_1) \cos(k_{z2} l_2).
\end{split}
\label{Eq:BlochDisp}
\end{equation}
Here, $k_{z1}^2 = \epsilon_1 \omega^2/c^2 - \kappa^2$, $k_{z2}^2 = k_{z0}^2$, and 
$P_\rs = 1$ and $P_\rp = \epsilon_1$ depending on the polarization state; $l_1$ and $l_2$ are the thicknesses of the two layers and $\gamma_\lambda$ is the Bloch {wave-vector component parallel to the optical axis. In the long-wavelength limit this dispersion relation
reproduces the EMT expressions in Eqs.~(\ref{Eq:DisperionOrdinary}) and (\ref{Eq:DisperionExtraOrdinary}).}

When determining the Bloch bands from Eq.~(\ref{Eq:BlochDisp}) and adding this information to the plots of the
transmission coefficients we find the result shown in Fig.~\ref{Fig:TransmissionCoefficient2}(a). From this figure
it can be seen that the surface modes lie outside the two hyperbolic Bloch bands in the region labeled as 'SM'.
For comparison we show the EMT result in Fig.~\ref{Fig:TransmissionCoefficient2}(d). Here,  
we model the mHMM as a $9\cdot\Lambda = 180\,{\rm nm}$ layer of an effective uniaxial material with permittivities described by 
Eqs.~(\ref{Eq:epsperp}) and (\ref{Eq:epsparallel}) with $f = 0.5$ on a SiC substrate. Obviously this result
is different from the exact results, since the surface modes in the 'SM' region are not included
in the EMT. A fact which has important consequences for LDOS and heat flux calculations based on EMT~\cite{TschikinEtAl2013,Biehs2}.
Furthermore, the hyperbolic bands predicted by EMT are extended to infinite values of $\kappa$, whereas the exact
results predict Bloch bands which are limited with respect to the value of $\kappa$, i.e.\ for each frequency there
is a maximum $\kappa$ for which one can find hyperbolic Bloch modes. Finally, it is clear that the EMT makes no difference
between the different kinds of configurations. In Fig.~\ref{Fig:TransmissionCoefficient2}(a) we have chosen SiC
as topmost layer for both media. We can of course also chose to make the calculations for our mHMM with 
Ge as topmost layers for both mHMM. Then we obtain the result shown in Fig.~\ref{Fig:TransmissionCoefficient2}(b). The
surface mode in the 'SM' region corresponding to the surface phonon polaritons of the two topmost SiC layers in    
Fig.~\ref{Fig:TransmissionCoefficient2}(a) is fully suppressed in this case, so that for this configuration mainly the hyperbolic
modes are contributing to the radiative heat flux for $d = 10\,{\rm nm}$. If we chose a mixed configuration, which means that for one
mHMM SiC is the topmost layer and for the other one Ge is the topmost layer, we find the result shown in 
Fig.~\ref{Fig:TransmissionCoefficient2}(c). In this case the transmission coefficient has still some traces of the 
surface mode contribution in the 'SM' region.

\subsection{Radiative Heat Flux}

In order to discuss the radiative heat flux, we can make use of the expressions derived in Sec.~III. For small temperature 
differences between the two considered media, i.e.\ we have $T_1 = T + \Delta T$ and $T_2 = T$ with $\Delta T \ll T$, it is 
more convenient to study the heat transfer coefficient (HTC) instead of the heat flux $\Phi$. The HTC is defined as
\begin{equation}
  h =   \int_0^\infty \!\!\!  \frac{\rd \omega}{2 \pi} \, \frac{\rd \Theta(\omega,T)}{\rd T}  H_\omega\\
\label{Eq:htc0}
\end{equation}
where the {reduced} spectral heat transfer coefficient (sHTC) is 
\begin{equation}
  H_\omega = \sum_{\lambda = \rs,\rp} \int \!\!\frac{\rd^2\kappa}{(2 \pi)^2} \, \mathcal{T}_\lambda (\omega,\boldsymbol{\kappa}; d)
\label{Eq:HTC0}
\end{equation}
With this definition the heat flux is itself given by $\Phi = h \Delta T$. Throughout the manuscript we set $T = 300\,{\rm K}$.

\begin{figure}
  \subfigure[$d = 100\,{\rm nm}$]{\epsfig{file = 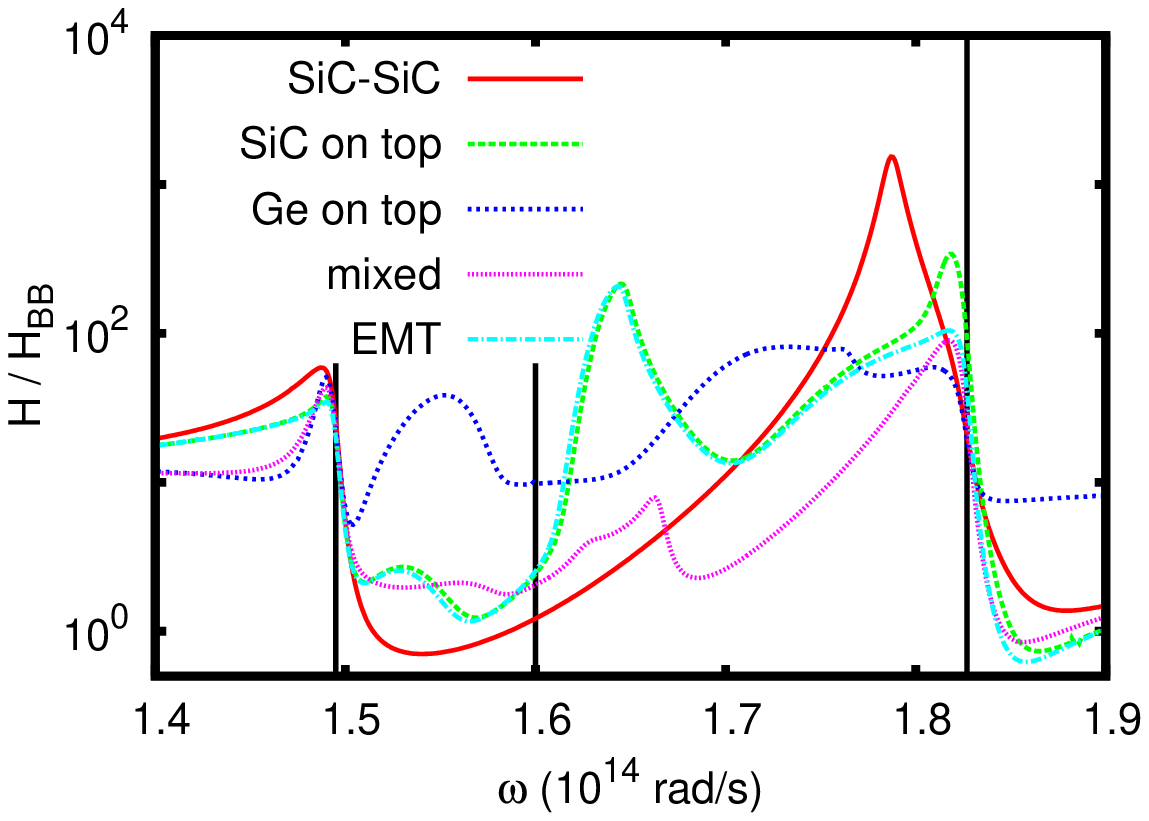, width = 0.45\textwidth}}
  \subfigure[$d = 10\,{\rm nm}$]{\epsfig{file = 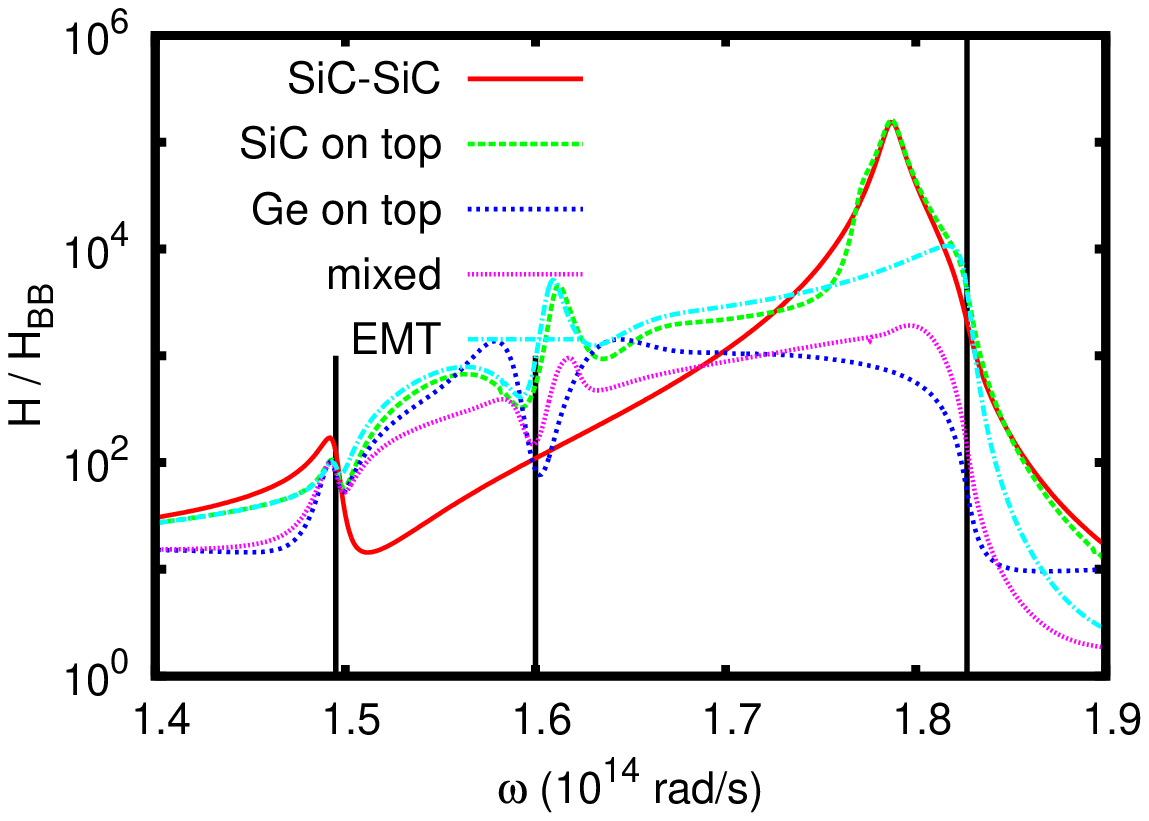, width = 0.45\textwidth}}
  \caption{Plot of the sHTC normalized to the blackbody result for different configurations and different vacuum gaps.
           For comparison the results of EMT are added. {The vertical lines mark the edges of the hyperbolic bands. }
           \label{Fig:sHTC}}
\end{figure}

In Fig.~\ref{Fig:sHTC} we first show the sHTC normalized to that of a blackbody for two SiC semi-infinite materials, 
two SiC/Ge mHMM with SiC as topmost layer, two SiC/Ge mHMM with Ge as topmost layer and finally the mixed case 
where one mHMM has SiC as topmost layer and the other one Ge as topmost layer. For comparison we have also included 
the results of EMT. In Fig.~\ref{Fig:sHTC}(a) we have $d = 100\,{\rm nm}$ and therefore $d \gg \Lambda$, 
since $\Lambda = 20\,{\rm nm}$. In this case, one could expect that EMT is a very good approximation. Indeed, the
EMT results are for most frequencies quite well reproducing the results of the case where SiC 
is the topmost layer. Nonetheless, the other two cases where Ge is the topmost layer or the mixed configuration are
not well described by EMT at all {which is due to the fact that we have used SiC as substrate for the
EMT calculation. But when Ge is on top then in the exact calculation we also have Ge as substrate. Therefore 
one needs to use a Ge substrate in the EMT calculation as well. We have checked that EMT gives a good approximation 
of the case where Ge is on top when using Ge as a substrate.} Furthermore, it can already 
be seen that for two bulk SiC media the radiative heat 
flux is mostly narrowband and concentrated around the surface-phonon polariton resonance 
at $\omega_{\rm SPhP} = 1.787\times10^{14}\,{\rm rad/s}$ whereas for the mHMM also hyperbolic modes contribute 
inside the reststrahlen band, so that for mHMM the heat flux is broadband as discussed in Refs.~\cite{Biehs2012,GuoEtAl2012}. 

For extremely small distances of $d = 10\,{\rm nm}$, we have $d < \Lambda$. In this case it cannot be expected that EMT gives
any reasonable result. By comparing EMT to the exact results in  Fig.~\ref{Fig:sHTC}(b) it can be seen that the EMT results
are actually not so far off from the exact results for most frequencies when SiC is topmost layer, but they lack of course the surface mode 
contribution of the first SiC layer as could already be observed in the transmission coefficients. This finding has 
important consequences for the LDOS and heat flux as discussed in Refs.~\cite{TschikinEtAl2013,Biehs2,LiuBright2014}. Compared to the mixed 
configuration and the configuration with Ge as topmost layer EMT overestimated the heat flux inside the reststrahlen 
band for most frequencies, whereas compared to the configuration with SiC as topmost layer EMT underestimates the heat 
flux especially around $\omega_{\rm SPhP}$.

\begin{figure}
  \epsfig{file = 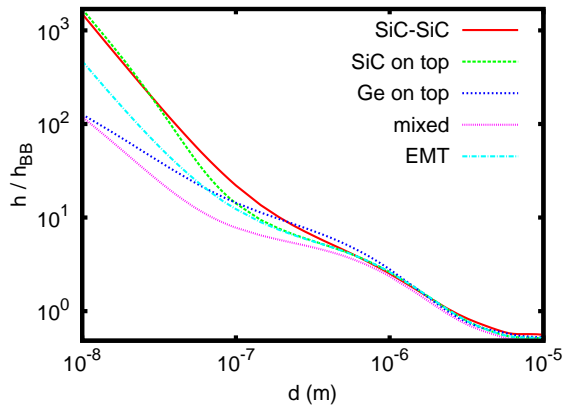, width = 0.45\textwidth}
  \caption{HTC normalized to the blackbody value $h_{\rm BB} = 6.12\,{\rm W}/({\rm m}^2 K)$ for two SiC half spaces,
           and two Ge/SiC mHMM with SiC as topmost layer, Ge as topmost layer and for the mixed case. In addition the result
           predicted by the EMT is plotted. 
           \label{Fig:HTC}}
\end{figure}

In Fig.~\ref{Fig:HTC} we show finally the HTC normalized to the blackbody value which is $h_{\rm BB} = 6.12\,{\rm W}/({\rm m}^2 K)$.
As can be expected from the sHTC the heat flux for the mHMM with SiC as topmost layer is slightly larger than
that of two SiC-SiC bulks for very small distances due to the extra contribution of the hyperbolic modes. The other 
configurations lead to much smaller heat fluxes for $d < 100\,{\rm nm}$. The EMT result 
underestimates the flux for the mHMM with SiC as topmost layer and it overestimates the heat flux of the 
other mHMM configurations in this distance regime. Nonetheless, the EMT result coincides with the exact values when SiC is on top 
for distances larger than $200\,{\rm nm} = 10 \cdot\Lambda$. Interestingly, in the intermediate distance regime between $100\,{\rm nm}$ and $1000\,{\rm nm}$
the mHMM with Ge as topmost layer gives larger heat fluxes than predicted by the EMT. Obviously, the overall heat flux level 
of mHMM is for most distances smaller than the heat flux between two simple SiC-SiC bulk materials. 

%

\subsection{Impact of the first single layer}

\begin{figure}
  \subfigure[$d = 100\,{\rm nm}$]{\epsfig{file = 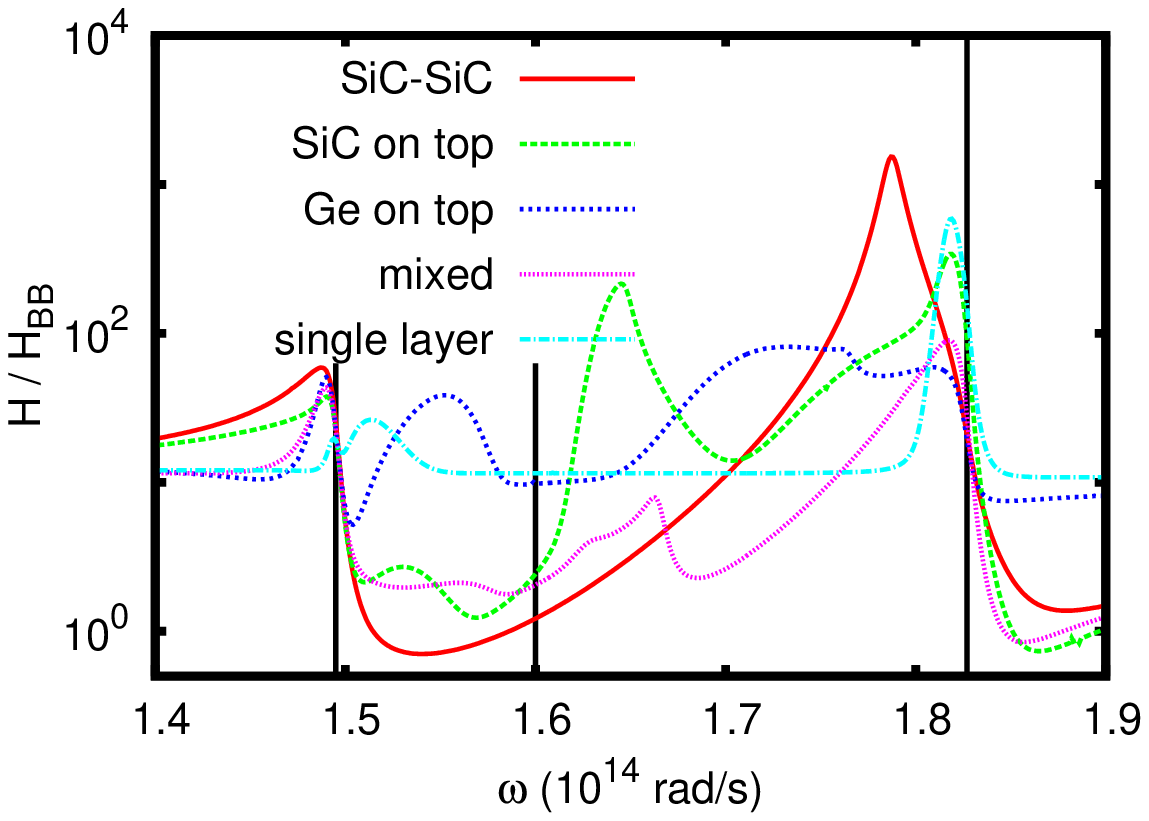 , width = 0.45\textwidth}}
  \subfigure[$d = 10\,{\rm nm}$]{\epsfig{file = 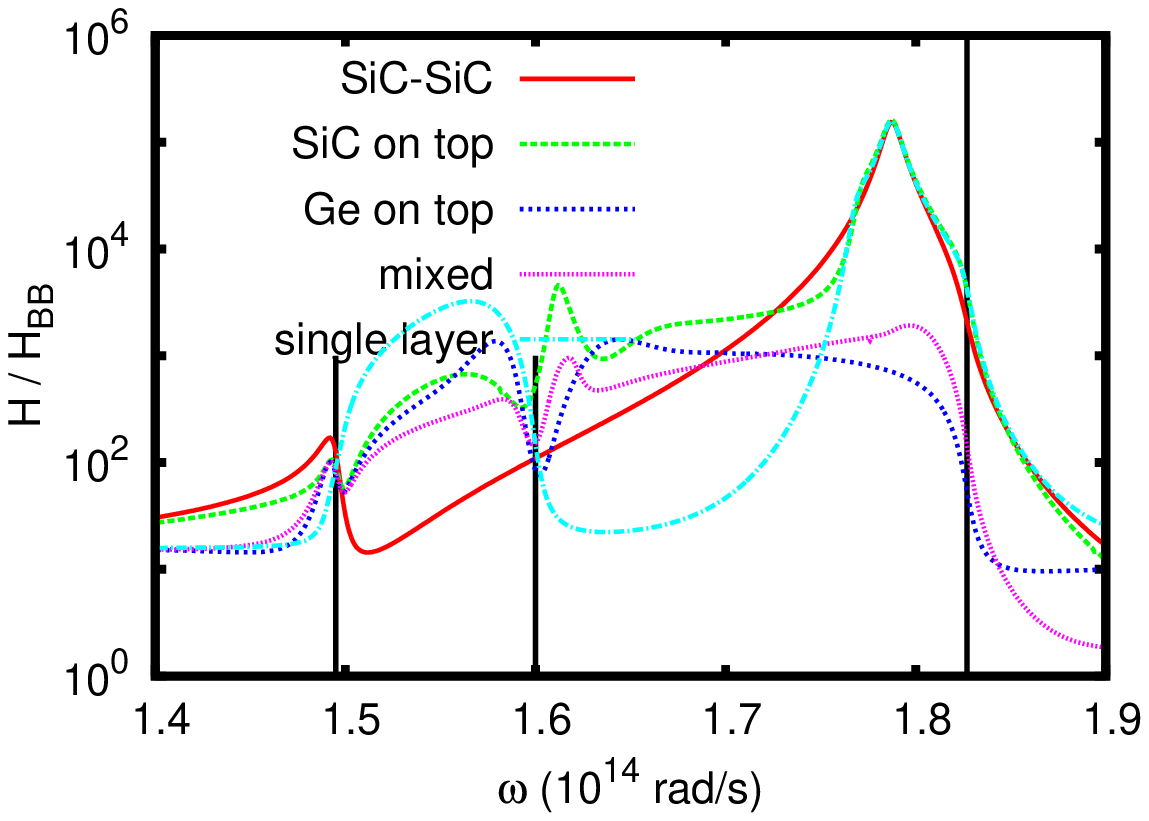 , width = 0.45\textwidth}}
  \caption{Plot of the sHTC as in Fig.~\ref{Fig:sHTC}, but instead of the EMT result we plot the values of the sHTC
           of two structures with a single 10nm SiC layer on a Ge substrate.
           \label{Fig:sHTC2}}
\end{figure}

Recently, the idea was brought forward that one can replace ``the HMM with a single thin layer, optimized for
even greater heat transfer''~\cite{MillerEtAl2014}. Therefore we want to compare the sHTC and the HTC
for our SiC/Ge mHMM structure with a structure consisting of a single 10nm layer of SiC on a semi-infinite 
Ge substrate. In Fig.~\ref{Fig:sHTC2} we show again the sHTC as in Fig.~\ref{Fig:sHTC} but this time we 
compare the results for the different configurations of the SiC/Ge mHMM structure with the results with a 
structure which is simply given by a single 10nm SiC slab on a Ge substrate. It can be seen in Fig.~\ref{Fig:sHTC} 
that the sHTC for the single layer structure is quite different from the sHTC of the mHMM no matter what configuration. 
The two peaks which can be seen close to the edges of the reststrahlenband are due to the coupled surface phonon 
polaritons of the thin single layer. 

\begin{figure}
  \epsfig{file = 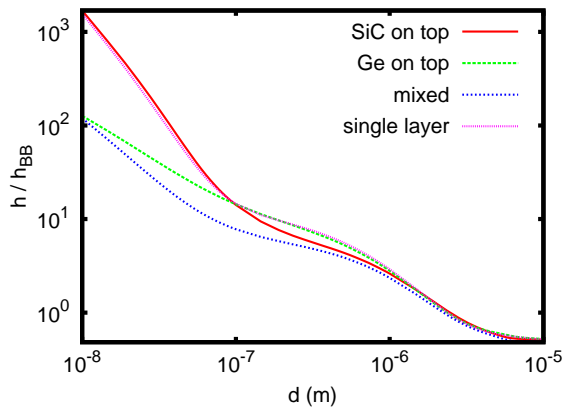, width = 0.45\textwidth}
  \caption{HTC normalized to the blackbody value $h_{\rm BB} = 6.12\,{\rm W}/({\rm m}^2 K)$ a SiC/Ge mHMM as in 
           Fig.~\ref{Fig:HTC} but with the HTC of two structures with a single 10nm SiC layer on a Ge substrate.
           \label{Fig:HTCsinglelayer}}
\end{figure}

When looking at the full HTC in Fig.~\ref{Fig:HTCsinglelayer} it seems that the HTC of the single layer structure
is interpolating that of the mHMM with Ge as topmost layer and that with SiC as topmost layer. For $d < 100\,{\rm nm}$
the HTC of the single layer resembles that of the mHMM with SiC on top and for $d > 100\,{\rm nm}$ it resembles
the case of the  mHMM with Ge on top. Therefore the statement of Ref.~\cite{MillerEtAl2014} seems to be correct
although in our plots the mHMM gives slightly larger results than the single layer structure for $d \ll 100\,{\rm nm}$. 

\begin{figure}
  \subfigure[mHMM vs.\ single layer]{\epsfig{file = 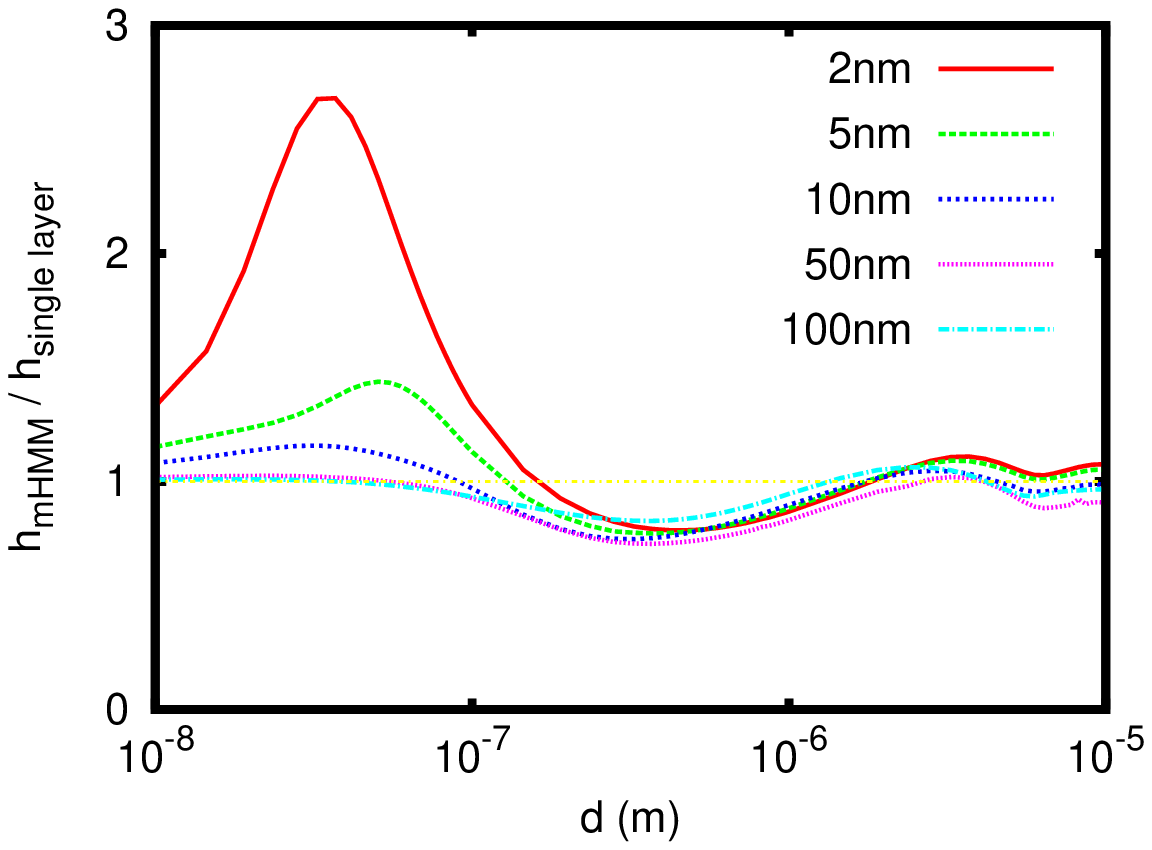, width = 0.45\textwidth}}
  \subfigure[mHMM vs.\ bulk SiC]{\epsfig{file = 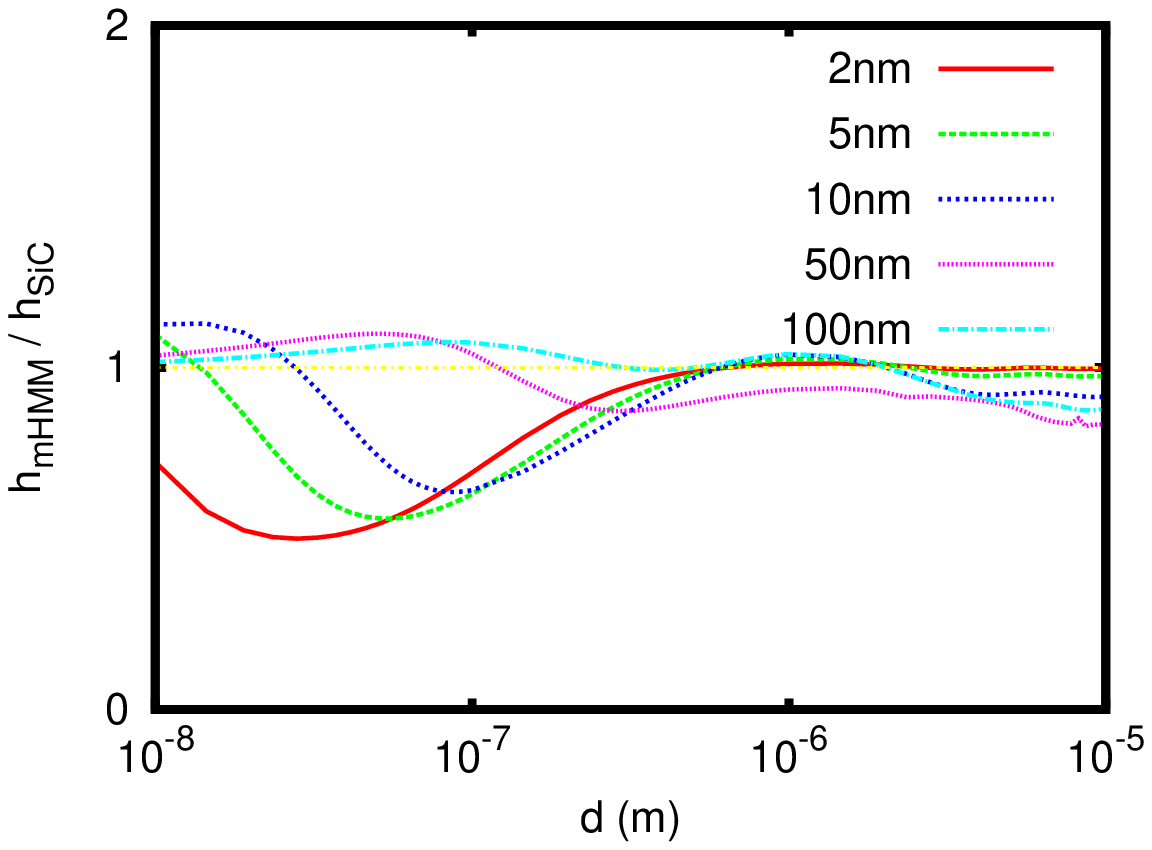, width = 0.45\textwidth}}
  \caption{HTC of a SiC/Ge mHMM with SiC as topmost layer with layer thicknesses of  $2\,{\rm nm}, 5\,{\rm nm}, 10\,{\rm nm}, 50\,{\rm nm}$, and
           $100\,{\rm nm}$ ($f = 0.5$ and the number of unit cells is $N = 9$) compared to the HTC of (a) a single SiC film of same thickness
           as the topmost SiC layer on a Ge substrate and (b) a SiC halfspace.
           \label{Fig:HTCsinglelayer2}}
\end{figure}

To test the statement for mHMM with different $\Lambda$ we have plotted in Fig.~\ref{Fig:HTCsinglelayer2}(a) the HTC of 
SiC/Ge mHMM with SiC as topmost layer with different layer thicknesses but with a fixed number of unit cells ($N = 9$) and with fixed $f = 0.5$ normalized to that of a single thin layer of SiC with the
same thickness as in the mHMM on a semi-infinite Ge substrate. It can be seen that for $d < 100\,{\rm nm}$ the HTC for the two mHMM tends
to be larger than that of the single SiC films in general. Only for relatively thick layers of $50\,{\rm nm}$ and $100\,{\rm nm}$
the HTC of the mHMM is the same as that of the single layer for $d < 100\,{\rm nm}$. This suggests that the above
statement is not precisely correct, but nevertheless one can say that when it is necessary to have large heat flux levels 
then mHMM seem to have no big advantage over a single film of phonon-polaritonic material. But the same is true if one compares the
HTC of the mHMM with that of bulk SiC as can be seen in Fig.~\ref{Fig:HTCsinglelayer2}(b). Nonetheless, with mHMM
one can have larger heat fluxes than for the phonon-polaritonic bulk material or a single film of that material in particular
when the period of the multilayer structure is small. We have checked that this statement remains true when Ge is the topmost layer.

\begin{figure}
  \epsfig{file = 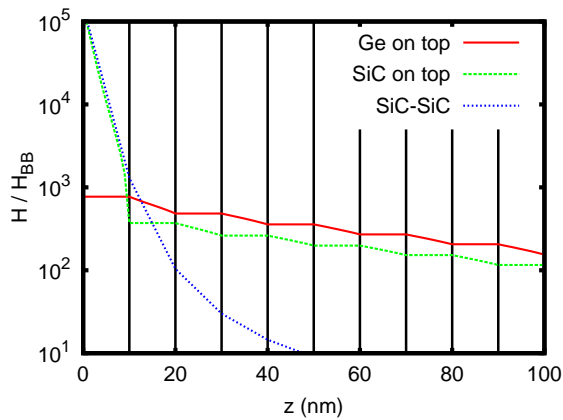, width = 0.45\textwidth}
  \caption{sHTC inside a receiver which is separated by a 10nm gap from a bulk SiC emitter at the surface phonon
           polariton resonance $\omega = \omega_{\rm SPhP} = 1.787\times10^{14}\,{\rm rad/s}$. The receivers are
           Bulk SiC and a SiC/Ge mHMM with $\Lambda = 10\,{\rm nm}$ and $f = 0.5$ for the case where SiC is topmost
           layer and Ge is topmost layer ($N = 9$). $z = 0$ is the interface of the receiver with the vacuum gap. The vertical
           lines mark the interfaces between the Ge and SiC layer inside the mHMM. It can be seen that the penetration
           depth inside the mHMM with Ge on top is much larger than in the two other cases where most of the incoming
           heat flux is dissipated inside the first layer or inside a region of about 10nm thickness.
           \label{Fig:HTCpenetration}}
\end{figure}

\subsection{Surface modes vs.\ hyperbolic modes}

One advantage of mHMM is that by chosing different orderings and filling fractions of the layers of the mHMM 
the frequency bands in which the material has a strong thermal emission can be controlled. Another more
striking advantage is that the hyperbolic modes are propagating inside the mHMM~\cite{Slawa2014,Tschikin2015}, whereas the surface modes 
are strongly confined to surface of the thermal emitter~\cite{BasuZhang2009,BasuZhang2011} as is also demonstrated in Fig.~\ref{Fig:HTCpenetration}. This property 
might be interesting for applications which involve the transport of heat radiation by a mHMM~\cite{MessinaEtAl2016} or 
applications which are based on light absorption within a thick layer close to the surface like near-field thermophotovoltaics~\cite{MatteoEtAl2001,NarayanaswamyChen2003,ParkEtAl2007,BernadiEtAl2015}, for instance. Therefore it is reasonable to quantify the
the contribution of the hyperbolic modes and/or Bloch modes to the heat flux. We can separate these contributions by 
using Eq.~(\ref{Eq:BlochDisp}) in order to distinguish between the Bloch modes and the surface modes of the mHMM structure.

\begin{figure*}
  \subfigure[SiC on top: total]{\epsfig{file = 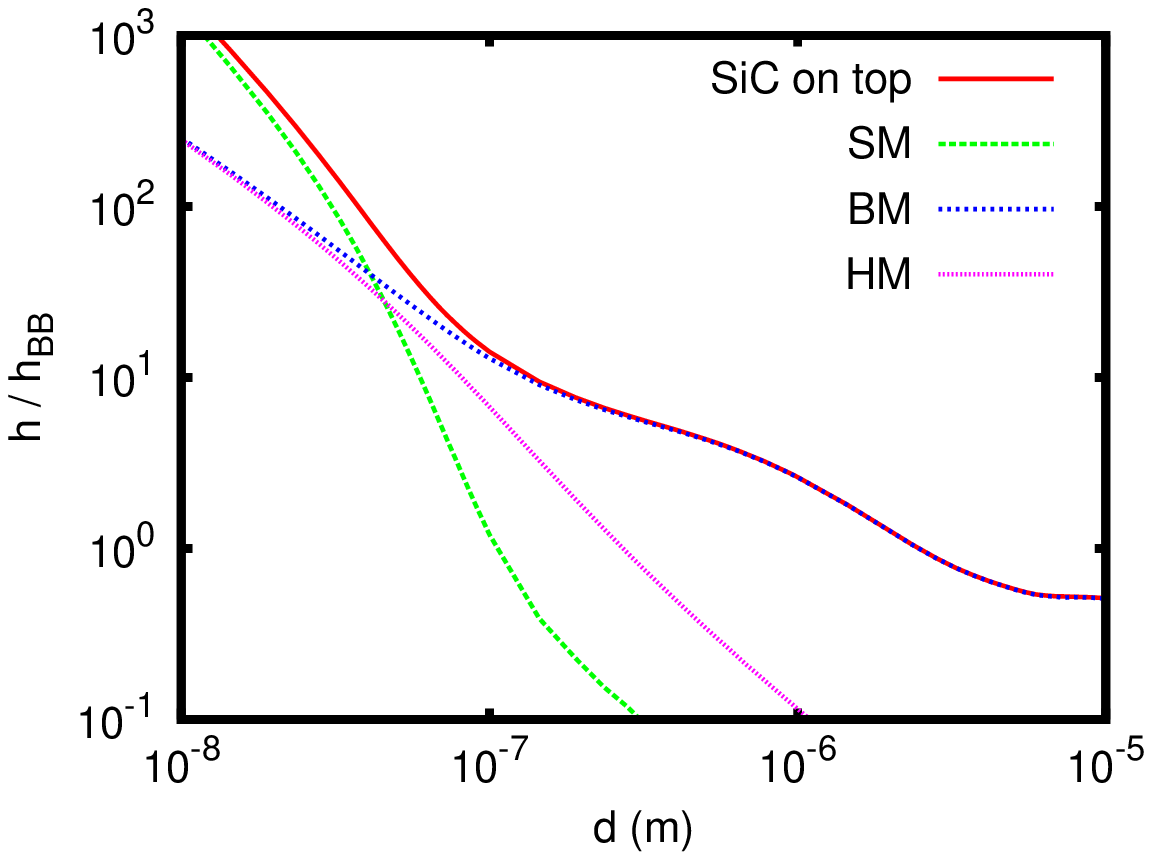, width = 0.45\textwidth}}
  \subfigure[SiC on top: relative]{\epsfig{file = 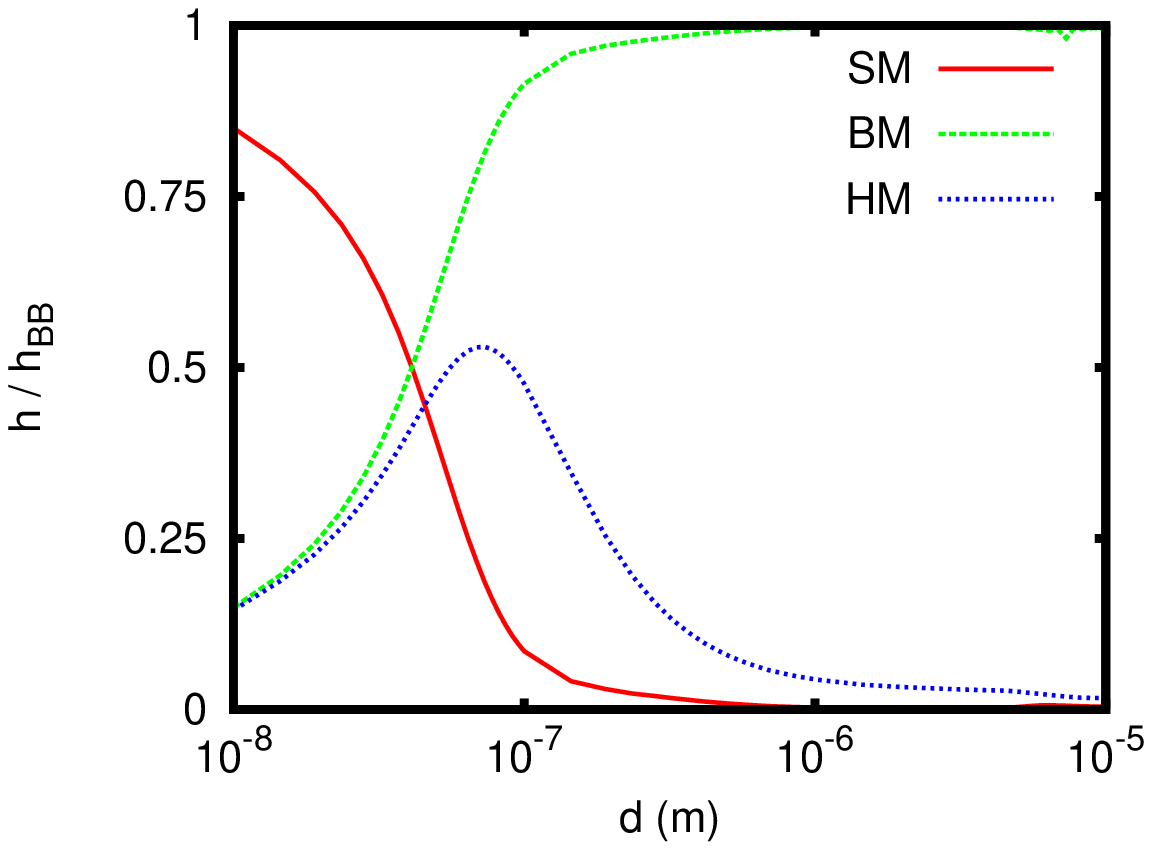, width = 0.45\textwidth}}
  \subfigure[SiC on top: total]{\epsfig{file = 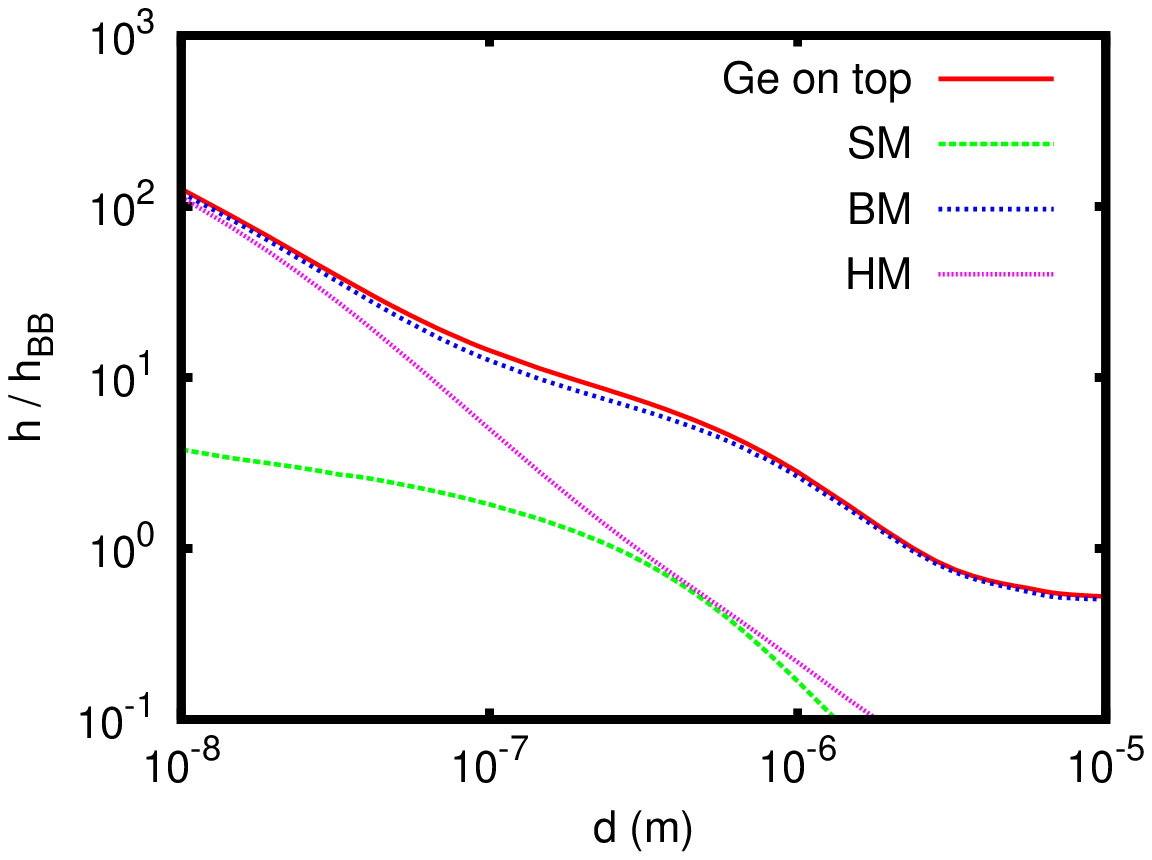, width = 0.45\textwidth}}
  \subfigure[SiC on top: relative]{\epsfig{file = 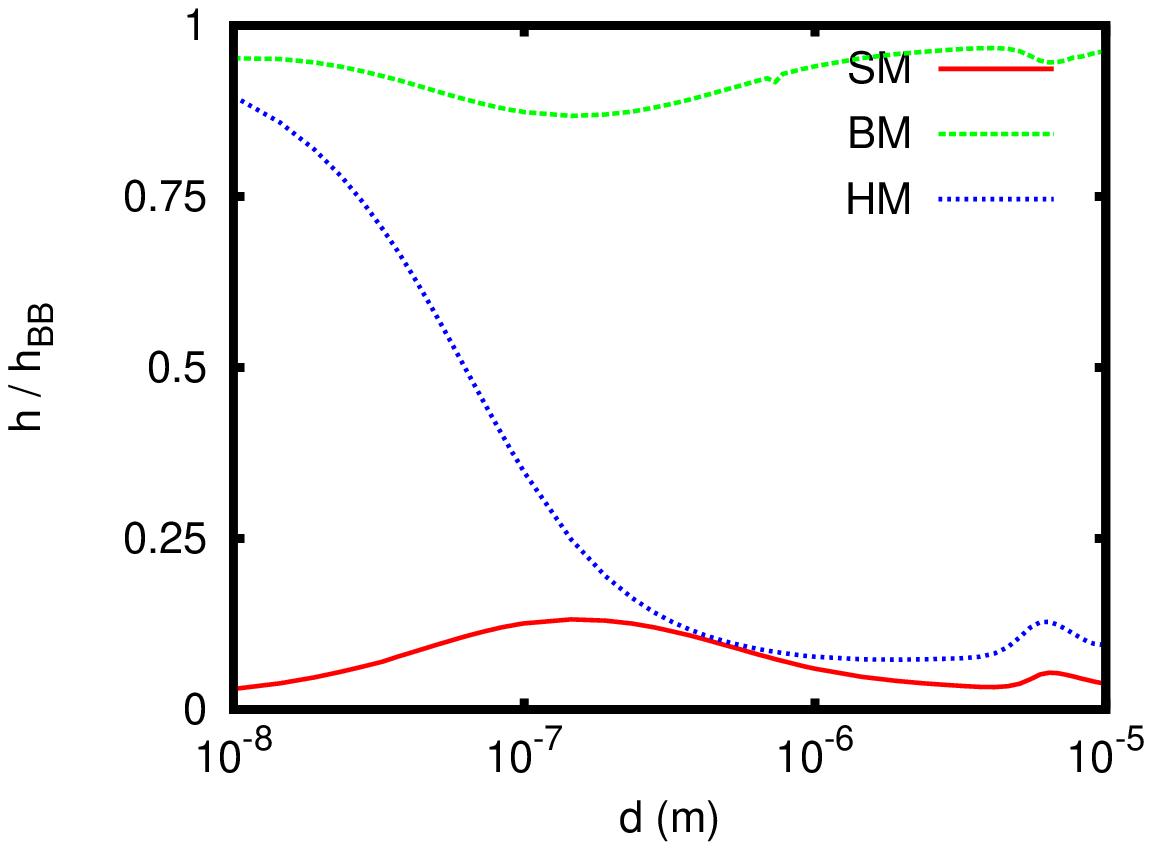, width = 0.45\textwidth}}

  \caption{Here we show (a),(c) the HTC due to the surface modes (SM), Bloch modes (BM), and hyperbolic modes (HM) in comparison to 
           the full HTC for the SiC/Ge mHMM for the two different configurations (SiC on top and Ge on top). 
           In (b),(d) we show the relative contributions of the SM, BM and HM.\label{Fig:HTC_SiContopSMBMHM}}
\end{figure*}

In Fig.~\ref{Fig:HTC_SiContopSMBMHM}(a) we show the different contributions. For distances larger than $d = 100\,{\rm nm}$ the
Bloch modes which are in this case mainly the 'normal' frustrated total internal reflection modes dominate the radiative heat flux.
Around $d = 100\,{\rm nm}$ the hyperbolic frustrated total internal reflection modes give the largest contributions of slightly more
than 50\% of the total heat flux as can be seen in  Fig.~\ref{Fig:HTC_SiContopSMBMHM}(b). For distances smaller than $d = 100\,{\rm nm}$
the surface modes start to dominate the radiative heat flux. At $d = 10\,{\rm nm}$ nearly 85\% of the heat flux is due to the surface
modes and nearly 15\% due to the hyperbolic modes. Note, that at $d = 10\,{\rm nm}$ the contribution of the hyperbolic modes is 
nonetheless 245 times larger than the blackbody value. If we choose Ge as topmost layer then we have already seen that the
surface mode contribution is suppressed. Consequently, we can see in Fig.~\ref{Fig:HTC_SiContopSMBMHM}(c) and (d)
that the heat flux is at $d = 10\,{\rm nm}$ nearly 90\% purely hyperbolic and the surface mode contribution is less than 4\%. The absolut
heat flux carried by the hyperbolic modes is still 114 times larger than that of a blackbody. These values can be improved by
chosen smaller periods of the mHMM. A similar result is obtained in the mixed case and therefore not shown here. Hence, if a mHMM 
is needed which allows for a large contribution of HM to the heat flux that means
if a true hyperbolic thermal emitter is needed then the configuration with Ge on top (with the material which does not itself support 
surface modes) is advantageous. The absolut heat flux values cannot be as high as for bulk SiC or as for a thin SiC layer for such a true
hyperbolic emitter, but the advantage is clearly that the hyperbolic modes are propagating in nature. Therefore the absorption and 
generation of heat is not limited to a thin layer at the surface of the emitter, but to a much broader region~\cite{Slawa2014,Tschikin2015}.

%
%
%

\section{Summary and Conclusion}

In conclusion, we have reviewed in large detail the properties of mHMM. In particular we have discussed the formation of hyperbolic bands,
the impact of the different configurations on the heat flux, the relative contribution of the hyperbolic heat flux channels to the full
heat flux, and we have compared the heat flux for mHMM with that of a single layer structure. We have seen that the mHMM emitters
are broadband in general and the full heat flux level is in most cases only slightly larger than that of a bulk medium or a single layer.
For a true mHMM emitter where the heat flux is dominated by hyperbolic modes (i.e.\ when Ge is topmost layer) the absolute heat flux level 
is typically inferior to bulk but not necessarily to single layer structures. Nonetheless, for mHMM with $\Lambda = 20\,{\rm nm}$ it can still easily be larger than 
100 times the blackbody value and for  $\Lambda = 4\,{\rm nm}$ it can be larger than 300 times the blackbody value for $d = 10\,{\rm nm}$. 
Hence, the advantage of mHMM is not necessarily the large heat flux level, but the fact that relatively large heat flux levels are combined 
with a large penetration depth of the heat flux, which is for mHMM by orders of magnitude larger than 
that of phonon-polaritonic materials as SiC~\cite{Slawa2014,Tschikin2015}. As a consequence the heat flux is not absorbed in an ultra-thin 
layer close to the surface~\cite{BasuZhang2009} but on a relatively thick layer. This property might be exploited in thermal management
applications, or for guiding near-field heat fluxes over far-field distances~\cite{MessinaEtAl2016} and also for near-field thermophotovoltaics 
where the conversion efficiency highly depends on the volume in which the light is absorbed~\cite{ParkEtAl2007,BernadiEtAl2015}. 

Other realizations of hyperbolic structures made with nanowires seem to have larger heat flux levels than mHMM and might even overcome the 
heat flux levels of the corresponding bulk materials from which the HM are made of by purely hyperbolic heat flux channels as shown in Refs.~\cite{Biehs2012,LiuShen2013,LiuEtAl2014HT,ChangEtAl2016}. It has been shown recently that this performance can even be improved drastically by covering HMM with graphene~\cite{LiuEtAl2014} so that even the theoretical limit derived in Refs.~\cite{Biehs2012,MillerEtAl2015} is reached. Most of these works on nanowire HMM are based on EMT, because a full numerical treatment is not as simple as for mHMM. However, a comparison of EMT with exact results for wHMM can be found in Ref.~\cite{Mirmoosa2014} but more exact calculations are necessary to check the validity of the EMT in near-field regime.

Up-to date only a few proposals of direct applications of HMs have been suggested. Apart from the waveguide for a long range transport of the near-field heat radiation~\cite{MessinaEtAl2016},  HMs could open a new avenue in the near-field energy conversion to conceive broadband emitters or receivers (PV cell). Here, further studies are needed which also take into account the impact of 
temperature gradients inside the materials~\cite{MessinaEtAl2016b}. Also, the singular properties of HMs could be exploited in far-field regime to design directionnal thermal emitters~\cite{Barbillon}, for instance.

%
%

\end{document}